\documentclass[11pt]{amsart}

\numberwithin{equation}{section} \oddsidemargin=-.0cm
\usepackage{amssymb}
\usepackage{amsmath}
\usepackage{color}
\usepackage{graphicx, graphics, subfigure}
\usepackage{amsfonts, amssymb, amsthm}

\usepackage{sidecap}
\usepackage{needspace} %used to avoid materials being split over a page break by placing e.g. \needspace{1\baselineskip} in the desired place

\sidecaptionvpos{figure}{c}

\usepackage[colorlinks=true, pdfstartview=FitV, linkcolor=blue,
            citecolor=blue, urlcolor=blue]{hyperref}  % the hyperref package should be the last package in order the link to 	
\usepackage[sort]{cite}
  \usepackage{paralist}
  \usepackage{graphics} %% add this and next lines if pictures should be in esp format
  \usepackage{epsfig} %For pictures: screened artwork should be set up with an 85 or 100 line screen
\usepackage{graphicx}  
\usepackage{epstopdf}%This is to transfer .eps figure to .pdf figure; please compile your paper using PDFLeTex or PDFTeXify.

\usepackage{epstopdf}
\usepackage{bbm} % for boldfaced numbers
\usepackage{amsfonts, amssymb, amsthm}
\usepackage{color}
\usepackage{needspace} %used to avoid materials being split over a page break by placing e.g. \needspace{1\baselineskip} in the desired place

\usepackage{booktabs} % for fine tuning tables

 \usepackage[colorlinks=true]{hyperref}
\hypersetup{urlcolor=blue, citecolor=blue}

\numberwithin{equation}{section}

\theoremstyle{definition}

\newcommand{\ep}{\varepsilon}

\graphicspath{{./Figures_Galerkin/},{../figures/}}
\newtheorem{thm}{Theorem}[section]
\newtheorem{lem}{Lemma}[section]
\newtheorem{defi}{Definition}[section]
\newtheorem{ex}{Example}[section]
\newtheorem{prop}[thm]{Proposition}

\newtheorem{rem}{Remark}[section]
\newtheorem{cor}{Corollary}[section]
\newenvironment{Proof}[1][Proof]{\begin{trivlist} \item[\hskip \labelsep {\it \bf  #1.}]}{\end{trivlist}}
\def\bpp{\begin{Proof}}
\def\epp{\end{Proof}}

\def\bt{\begin{thm}}
\def\et{\end{thm}}

\def\bex{\begin{ex}}
\def\eex{\end{ex}}

\def\bl{\begin{lem}}
\def\el{\end{lem}}
\def\bd{\begin{defi}}
\def\ed{\end{defi}}
\def\bc{\begin{cor}}
\def\ec{\end{cor}}
\def\bp{\begin{proof}}
\def\ep{\end{proof}}
\def\br{\begin{rem}}
\def\er{\end{rem}}

\def\bprop{\begin{prop}}
\def\eprop{\end{prop}}

\def\Forall{\text{ } \forall \:}
\def\d{\mathrm{d}}

\def\be{\begin{equation}}
\def\ee{\end{equation}}
\def\bes{\begin{equation*}}
\def\ees{\end{equation*}}
\def\bea{\begin{equation} \begin{aligned}}
\def\eea{\end{aligned} \end{equation}}
\def\beas{\begin{equation*} \begin{aligned}}
\def\eeas{\end{aligned} \end{equation*}}

\def\bi{\begin{itemize}}
\def\ei{\end{itemize}}
\def\ben{\begin{enumerate}}
\def\een{\end{enumerate}}

\def\tr{\mathrm{tr}}
\renewcommand{\boldsymbol}[1]{{#1}} 
\def\span{\mathrm{span}}

\definecolor{greenrb}{rgb}{0.2,0.6,0.2}
\definecolor{rred}{rgb}{0.7,0,0.1}

\newcommand{\mk}{\color{black}}

\newcommand{\mg}{\color{black}}

\usepackage[normalem]{ulem}

\title[Galerkin Approximations of Nonlinear Delay Differential Equations] %Use the shortened version of the full title
{Low-Dimensional Galerkin Approximations of Nonlinear Delay Differential Equations}

\author[M. D. Chekroun]{Micka\"el D. Chekroun}
\address[MDC]{Department of Atmospheric \& Oceanic Sciences, University of California, Los Angeles, CA 90095-1565, USA} 
\email{mchekroun@atmos.ucla.edu}

\author[M. Ghil]{Michael Ghil}
\address[MG]{Geosciences Department and Laboratoire de M\'et\'eorologie Dynamique (CNRS and IPSL),\'Ecole Normale Sup\'erieure, F-75231 Paris Cedex 05, France; and Department of Atmospheric \& Oceanic Sciences, University of California,Los Angeles, CA 90095-1565, USA}
\email{ghil@atmos.ucla.edu}

\author[H. Liu]{Honghu Liu}
\address[HL]{Department of Atmospheric \& Oceanic Sciences, University of California, Los Angeles, CA 90095-1565, USA; and Department of Mathematics, Virginia Polytechnic Institute and State University, Blacksburg, Virginia 24061, USA}
\email{hhliu@vt.edu}

\author[S. Wang]{Shouhong Wang}
\address[SW]{Department of Mathematics, Indiana University}
\email{showang@indiana.edu}



\begin{document}

\begin{abstract}
This article revisits the approximation  problem of systems  of nonlinear delay differential equations (DDEs) 
by a set of ordinary differential equations (ODEs).   
We work in Hilbert spaces endowed with a natural inner product including a point mass, 
and introduce polynomials orthogonal with respect to such an inner product that live in the domain of 
the linear operator associated with the underlying DDE. These 
polynomials are then used to design a general Galerkin scheme for which 
we derive rigorous convergence results and show that it can be  
numerically implemented via simple analytic formulas.   
The scheme so obtained is applied to three nonlinear DDEs, two autonomous and one forced:
(i) a simple DDE with distributed delays whose solutions recall Brownian motion;  
(ii) a DDE with a discrete delay that exhibits bimodal and chaotic dynamics; 
and (iii) a periodically forced DDE with two discrete delays arising in climate dynamics. 
In all three cases, the Galerkin scheme introduced in this article 
provides a good approximation by low-dimensional ODE systems 
of the DDE's strange attractor, as well as of the statistical features that characterize its nonlinear dynamics.

\end{abstract}
\maketitle

\section{Introductuion}

Systems of delay differential equations (DDEs) are widely used in many fields such as the biosciences,
climate dynamics, chemistry, control theory, economics, and engineering 
\cite{Bhattacharya_al82, Diaz2014, Diekmann_al95, GCStep15, Ghil_Childress'87, GZT08, Hale_Lunel93, Sieber2014, Kuang93, LS10, MacDonald89, Michiels_al07, Roques_al15, Smith11, Stepan89}.  In particular, certain DDEs or more general differential equations with retarded arguments can be derived from hyperbolic partial differential equations that support wave propagation \cite{chekroun_glatt-holtz,Galanti_al00,Hale_Lunel93}.

In contrast to ordinary differential equations (ODEs), the state space associated even with a scalar DDE is infinite-dimensional, due to the presence of time-delay terms, which require providing initial data over an interval $ [-\tau,0],$ where $\tau >0$ is the delay. 
It is often desirable, though, to have low-dimensional ODE systems that capture qualitative features, 
as well as approximating certain quantitative aspects of the DDE dynamics. 

The derivation of ODE approximations of DDEs involves, in general, two types of function spaces as state space: 
 that of continuous functions $C([-\tau, 0]; \mathbb{R}^d)$, and the Hilbert space $L^2([-\tau, 0); \mathbb{R}^d)$. 
{\mg The former spaces have}  been extensively used in the case of bifurcation analysis \cite{Casal_al80,Chow_al77,Das_al02,Faria_al95,Kazarinoff_al78,Nayfeh08,wischert1994delay}, {\mg while the latter are} typically adopted in situations where quantitative accuracy is an important factor, such as in optimal control \cite{Banks_al78,kappel1978autonomous,Banks_al84,Kappel86,Kappel_al87,Kunisch82,Ito_Teglas86,banks1979spline}. 

Within the Hilbert {\mg space} setting, different basis functions have been proposed to decompose  
the state space; {\mg these} include, among others, step functions \cite{Banks_al78,kappel1978autonomous}, splines \cite{banks1979spline,Banks_al84}, and orthogonal  polynomial functions, such as Legendre polynomials \cite{Kappel86,Ito_Teglas86}. Compared with step functions or splines, the use of orthogonal polynomials leads typically to ODE approximations with lower dimensions, for a given precision \cite{banks1979spline,Ito_Teglas86}. On the other hand, classical polynomial basis functions 
do not live in the domain of the linear operator underlying the DDE, which leads to technical complications in establishing convergence results \cite{Kappel86,Ito_Teglas86}; see Remark~\ref{Rmk_problems_to_overcome}(iii) below.

In the present article, we propose to avoid these technical difficulties in 
approximating DDEs as systems of ODEs by using an alternative polynomial basis:  the elements of this basis belong naturally to the domain of the underlying linear operator, but  they have not been used in the DDE literature so far. The polynomials we shall use are named after  Koornwinder~\cite{Koo84}, who investigated polynomials that are orthogonal with respect to weight functions adjoining point masses, as discussed in Section~\ref{sect:basis} below. This polynomial basis turns out to be particularly useful not only for the rigorous analysis of polynomial-based Galerkin approximations of nonlinear systems of DDEs,  as shown in Section~\ref{Sec_Galerkin_approx}, but also for their numerical treatment, cf. Section~\ref{Sect_Numerics}.

Useful new properties of the Koornwinder polynomials are identified in Lemma ~\ref{Fundamental_lemma} for the scalar case, and a generalization of these polynomials to the vector case is given in Section~\ref{Sec_Vectorization}; the latter includes 
the multi-dimensional extension of  Lemma ~\ref{Fundamental_lemma}, namely Lemma \ref{Super_Fundamental_lemma}.
We show that  these properties are essential for checking key stability and convergence conditions in {\mg Lemmas~\ref{Lem_A2}  and \ref{Lem_A1}. Standard Galerkin approximation results for abstract nonlinear ODEs} in Hilbert spaces are recalled in Theorem \ref{ParisVI_thm} and the rest of Section~\ref{Subsect_ODE_Galerkin}. They are then applied, with the help of Lemmas~\ref{Lem_A2}  and \ref{Lem_A1}, to  nonlinear systems of DDEs in Section~\ref{Subsect_DDE_Galerkin}.

Finite-time uniform convergence results are then derived for the proposed Galerkin approximations of nonlinear systems of DDEs, subject to simple and checkable conditions on the nonlinear term. 
These conditions are identified in Section~\ref{Subsect_DDE_Galerkin};
see Corollaries~\ref{Cor_DDE_local_Lip_Case1} and \ref{Cor_DDE_local_Lip_Case2}. 
The results apply to a  broad class of nonlinear systems of  DDEs, as discussed in Section~\ref{Sec_examples}. 

The proposed framework yields a simple numerical calculation of the corresponding Galerkin approximations. Their coefficients 
are easily computable from the original system of DDEs by relying on simple recurrence formulas, cf.~Proposition~\ref{thm:Pn}, and by solving upper triangular systems of linear equations, cf.~Proposition~\ref{prop:dPn}; see Section \ref{Sec_Galerkin_analytic} and Appendix~\ref{Appendix_systems}.

Finally, we outline here a useful
interpretation of our proposed scheme regarding the finite-dimensional  approximation of the linear part $\mathcal{A}$ of general systems of DDEs, when considered in the framework of Hilbert spaces, cf.~\eqref{Def_A2}. This interpretation relies on a formulation of the evolution in time of the  
initial state $\{x(\theta): \theta \in [-\tau, 0]\}$ as the solution of a partial differential equation (PDE);
see also Remark \ref{PDE_rem}. 

 To do so, we first distinguish between the {\it historic part} of the evolving state, $\{x(t + \theta): \theta \in [-\tau, 0)\}$, and the {\it state part}, $\{x(t)\}$. Denoting by $u(t,\theta)$ the historic part, one can  then rewrite, for instance, the simple linear DDE 
\be\label{lin_case}
\dot{x}=x(t-\tau), \quad \tau>0,
\ee
as the linear PDE
\be\label{lin_PDE}
\partial_t u = \partial_{\theta}u, \quad -\tau \le \theta < 0,
\ee
with the boundary condition
\be\label{PDE_BC}
\partial_{t} u|_{\theta = 0} = u(t,-\tau), \;\; t \geq 0.
\ee

The key point is that, roughly speaking, the local differential operator $v\mapsto \partial_{\theta} v$ --- obtained as the history component of $\mathcal{A}$, and written out explicitly in~Eq.~\eqref{Def_A}, for instance ---
is approximated here by the  nonlocal differential operator 
\be\label{nonlocal_PDE_intro}
v \mapsto \partial_{\theta} v +b_N(\theta)\Big(v(-\tau)- \partial_{\theta} v\big\vert_{\theta=0}\Big), \quad \theta \in[-\tau,0),
\ee
and that $b_N(\theta)$ --- expressed by means of Koornwinder polynomials, cf.~\eqref{bN-coeff} --- is a bounded oscillatory coefficient that vanishes in $L^2$ as $N\rightarrow \infty$; see  Lemma \ref{Fundamental_lemma}.  This nonlocal operator is the PDE representation for the history component of our Koornwinder-based Galerkin approximation $\mathcal{A}_N$ given in \eqref{Eq_AN}.

Note that terms such as $v(-\tau) - \partial_{\theta} v\big\vert_{\theta=0}$,  which is responsible  for the nonlocal aspect of 
\eqref{nonlocal_PDE_intro}, play an important role in the theory of numerical approximation of DDEs; see, for instance,\cite[Appendix A]{GZT08} and references therein.  In particular, this term provides the exact value of the jump  
associated with the boundary condition \eqref{PDE_BC}.
The first such jump occurs at $t=0$ in the derivative of solutions of Eq.~\eqref{lin_case} that emanate from a constant history\footnote{\
For instance, if $\tau =1$ and the history 
 is given by $\{\psi(\theta)\equiv c, \; - \tau  \le \theta < 0\}$, then the solution $x(t)$ to Eq.~\eqref{lin_case} is equal to $c(t+1)$ on $[0,1]$. This discontinuity leads to a jump $\psi(-1)-\partial_{\theta} \psi\big\vert_{\theta=0} = 
c$  in its time derivative at $t=0$.};  this discontinuity propagates to higher-order derivatives at subsequent, 
 integer multiples of the delay, $t=k\tau$, $k\in \mathbb{Z}^+$. 

The fact that this jump term is weighted by a vanishing term 
 suggests that, for a given degree of accuracy, good approximation can be expected
when using relatively low-dimensional Koornwinder-based approximations, as long as $b_N$ vanishes sufficiently quickly.  We do not address such numerical considerations here; see, however, Table \ref{table} in Remark \ref{PDE_rem} for results in the case of Eq.~\eqref{lin_case}. 

Instead, in Section~\ref{Sect_Numerics}, we 
provide several applications that show the proposed approximation to be not only rigorously justified, but very effective  
in nonlinear cases that yield quasi-periodic and chaotic, as well as nearly Brownian dynamics. In each case, low-dimensional 
 ODE systems succeed in approximating important topological as well as statistical features of the corresponding DDE's  nonlinear dynamics.

The article is organized as follows. In Section~\ref{sect:preliminaries}, we introduce the functional framework that will be adopted in Section \ref{Subsect_DDE_Galerkin} to recast a system of nonlinear DDEs into an abstract ODE. This framework relies on Hilbert spaces endowed with a natural inner product with a point mass.  Koornwinder polynomials are then introduced in Section~\ref{sect:basis}.  The convergence of the Galerkin ODE systems built  by projecting onto these polynomials to the original DDEs is proven in Section \ref{Sec_Galerkin_approx}. 

We provide explicit expressions of the Galerkin approximation 
in Section~\ref{Sec_Galerkin_analytic} for the scalar case,  and in Appendix~\ref{Appendix_systems} for nonlinear systems of DDEs. Finally, numerical applications to three nonlinear DDEs 
are provided in Section~\ref{Sect_Numerics}. These applications involve: (i) a simple DDE with distributed delays whose solutions recall Brownian motion \cite{sprott2007simple}; (ii) a DDE with a discrete delay that
 illustrates bimodal, as well as chaotic dynamics \cite{sprott2007simple}; and (iii) a periodically forced DDE with two discrete delays as a highly idealized model of the El Ni\~no-Southern Oscillation (ENSO: \cite[and references therein]{GZT08}).  In all three
examples, it is shown that our Galerkin scheme  provides a good approximation 
by low-dimensional ODE systems of the DDE's strange attractor, as well as the statistical features that characterize the associated nonlinear dynamics.

\section{Background and motivation} \label{sect:preliminaries}
We introduce in this section the functional framework that will be adopted in Section \ref{Subsect_DDE_Galerkin} for the derivation of Galerkin approximations of a given nonlinear system of DDEs.   Several function spaces can be used as a state space for the reformulation of a system of DDEs into an abstract ODE where among the most standard ones, those built-up out of the space of continuous functions on the interval $[-\tau,0]$ play an important role in the DDE theory; e.g.~\cite{Diekmann_al95,Hale_Lunel93}. 

In this article, we adopt instead the use of Hilbert  spaces which are more classically used in control or approximation theory of DDEs; see e.g., \cite{Banks_al78,burns1983linear,curtain1995,Kappel86,kappel1986equivalence,Kappel_al87,nakagiri1989controllability}. 
For a didactic expository of the associated theory of semigroups for (linear) systems of DDEs in this functional setting we refer to \cite[Sect.~2.4]{curtain1995}; see also \cite{burns1983linear}.

More precisely, the following Hilbert product space 
\be \label{H_space}
\mathcal{H} := L^2([-\tau,0); \mathbb{R}^d) \times  \mathbb{R}^d,
\ee 
will serve as our state space, and  will be endowed with the inner product defined for any  $(f_1, \gamma_1),\, (f_2, \gamma_2) \in \mathcal{H}$, as:
\be \label{H_inner}
 \langle (f_1, \gamma_1), (f_2, \gamma_2) \rangle_{\mathcal{H}} := \frac{1}{\tau} \int_{-\tau}^0\langle f_1(\theta), f_2(\theta) \rangle \d \theta  + \langle \gamma_1,\gamma_2\rangle,
\ee
where $\langle \cdot, \cdot \rangle$ denotes the Euclidean inner product of  $ \mathbb{R}^d$.

We will also make use of the following subspace of $\mathcal{H}$:
\be
\mathcal{V} := H^1([-\tau,0); \mathbb{R}^d) \times  \mathbb{R}^d,
\ee
where $H^1([-\tau,0); \mathbb{R}^d)$ denotes the standard Sobolev subspace of $L^2([-\tau,0); \mathbb{R}^d)$; see, e.g.~ \cite[Chap.~8]{brezis_book}. This space consists of functions that are square integrable and whose first-order weak derivatives exist in a distributional sense and are also square integrable. 

Instead of presenting the general nonlinear systems of DDEs considered in this article (see Section \ref{Subsect_DDE_Galerkin}), we introduce below a class of  scalar DDEs that will serve us to identify within a simple context,  the issues inherent to the Galerkin approximation of DDEs;  see Remark \ref{Rmk_problems_to_overcome} hereafter.

\bex\label{Ex_DDE_into_ODE}
In this example, we recall how a scalar DDE can be recast  into an abstract  ODE. 
For simplicity, we will focus on the following autonomous scalar DDE ($d=1$):
\be \label{Eq_DDE}
\frac{\d x(t)}{\d t} = a x(t) + bx(t-\tau) + c \int_{t-\tau}^t x(s)\d s + F\Big(x(t), \int_{t-\tau}^t x(s) \d s \Big),
\ee
where $a$, $b$ and $c$ are real numbers, $\tau> 0$ is the delay parameter, and $F$
is a given scalar nonlinear function. The case of nonlinear systems of DDEs will be dealt with in Section \ref{Subsect_DDE_Galerkin}.

The reformulation of Eq.~\eqref{Eq_DDE} into an abstract ODE is classical and proceeds as follows.  Let us denote by $x_t$ the time evolution of the history segments of a solution to Eq.~\eqref{Eq_DDE}, namely 
\be \label{shift}
 x_t(\theta):=x(t+\theta), \qquad t \ge 0, \qquad \theta \in [-\tau, 0].
\ee
Now, by introducing the new variable
\be
u(t) := (x_t,x(t))=(x_t, x_t(0)), 
\ee
Eq.~\eqref{Eq_DDE} can be rewritten as the following  abstract ODE: 
\be \label{eq:DDE_abs}
\frac{\d u}{\d t} = \mathcal{A} u + \mathcal{F}(u),
\ee
where the linear operator $\mathcal{A} \colon D(\mathcal{A}) \subset \mathcal{V} \rightarrow \mathcal{H}$ is defined by 
\bea \label{Def_A}
\lbrack \mathcal{A} \Psi \rbrack (\theta) & := \begin{cases}
{\displaystyle \frac{\d^+ \Psi^D}{\d \theta}}, &  \theta \in[-\tau, 0),  \vspace{0.4em}\\ 
{\displaystyle a \Psi^S + b\Psi^D(-\tau) + c \int_{-\tau}^0 \Psi^D(s)\d s}, & \theta = 0,
\end{cases} 
\eea
with the domain $\mathcal{A}$ given by (cf. \cite[Prop.~2.6]{Kappel86})
\be \label{D_of_A}
D(\mathcal{A}) = \Big \{(\Psi^D, \Psi^S) \in L^2([-\tau, 0); \mathbb{R})\times \mathbb{R} : \Psi^D \in H^1([-\tau, 0); \mathbb{R}), \lim_{\theta \rightarrow 0^-} \Psi^D(\theta) = \Psi^S 
\Big \};
\ee
and with the nonlinearity $\mathcal{F} \colon \mathcal{H} \rightarrow \mathcal{H}$ defined by 
\bea \label{Def_F}
[\mathcal{F} (\Psi) ](\theta) & := \begin{cases}
0, &  \theta \in[-\tau, 0),   \vspace{0.4em}\\ 
F \Big(\Psi^S, \int_{-\tau}^0 \Psi^D(s) \d s \Big), & \theta = 0, 
\end{cases}  \quad \Forall \Psi = (\Psi^D, \Psi^S) \in  \mathcal{H}.
\eea

With $D(\mathcal{A})$ such as given in \eqref{D_of_A}, the operator $\mathcal{A}$ generates a linear $C_0$-semigroup on $\mathcal{H}$   so that the Cauchy problem associated with the linear  equation  $\dot{u}=\mathcal{A} u$ is well-posed in the Hadamard's sense; see e.g~\cite[Thm.~2.4.6]{curtain1995}. The well-posedness problem for the nonlinear equation depends obviously on the nonlinear term $\mathcal{F}$ and we refer to Section \ref{Subsect_DDE_Galerkin} for a solution to  this problem within our functional framework; see also \cite{Webb76}.

\eex

\needspace{1\baselineskip}
\br\label{Rmk_problems_to_overcome}
\hspace*{2em}  \vspace*{-0.4em}
\bi
\item[{\mg (i)}] It is important to note that when instead of $L^2([-\tau, 0); \mathbb{R}^d)$, the space of continuous functions $X=C([-\tau, 0]; \mathbb{R}^d)$ endowed with the supremum norm  is retained  \cite{Hale_Lunel93}, the continuity requirement at $0$ in \eqref{D_of_A} is naturally satisfied. On the other hand,  $X$ is not a Hilbert space and the analysis of the {\it adjoint eigenvalue problem} \cite[Sect.~7.5]{Hale_Lunel93} is required for the derivation of low-dimensional ODE systems  which no longer contain memory terms \cite{wischert1994delay}. By working within the framework of  Hilbert spaces we avoid technicalities inherent to the analysis of  this adjoint problem. 

\item[(ii)]   When we consider the Hilbert space $\mathcal{H}$, a natural choice of set of  functions to decompose the solutions of \eqref{eq:DDE_abs} is constituted by the eigenfunctions of the operator $\mathcal{A}$ with domain $D(\mathcal{A})$.  When $\mathcal{A}$ does not contain distributed delay terms, these eigenfunctions are well-known and can be found in e.g.~\cite[Thm.~2.4.6]{curtain1995}. In case where the eigenvalues of $\mathcal{A}$ are all simple, this set of eigenfunctions  actually correspond to the set $\mathcal{E}$ of eigenfunctions in $C([-\tau, 0]; \mathbb{R}^d)$  \cite[Thm.~4.2, p.~207]{Hale_Lunel93}.  The latter set may fail however in approximating continuous functions \cite[Cor.~8.1, p. 222]{Hale_Lunel93} and can be even finite-dimensional \cite[p.~220]{Hale_Lunel93} which limits seriously its usage in practice\footnote{for the purpose of low-dimensional approximations.} if for instance snippets of solutions to  Eq.~\eqref{eq:DDE_abs} are spanned by elements outside of $\mathcal{E}$.\footnote{See however \cite[Thm.~2.5.10]{curtain1995} for a sufficient condition for the set of (generalized) eigenfunctions to be dense in $\mathcal{H}$ still for the case when $\mathcal{A}$ does not contain distributed delay terms.}

\item[(iii)] Due to the aforementioned limitations of the eigenfunctions, other basis functions are often used for the derivation of ODE systems to approximate the dynamics of the underlying DDE. Choices proposed in the literature include step functions \cite{Banks_al78,kappel1978autonomous}, splines \cite{banks1979spline,Banks_al84}, and orthogonal  polynomial functions such as Legendre polynomials \cite{Kappel86,Ito_Teglas86}.\footnote{It is also worth mentioning the more recent works \cite{Vyasarayani12,Wahi_al05}, in which interesting approximation schemes based on linear and  sine functions have been proposed  for the case of state dependent delays, and for which successful numerical performances have been reported although rigorous convergence results seem still to be lacking, within this approach.}

In most of the cases, a version of the Trotter-Kato theorem (see e.g.~\cite[Thm.~4.5, p.~88]{Pazy83}) is typically used to obtain finite-time uniform approximation results of the semigroup generated by $\mathcal{A}$. 
In the cases of step functions and splines, the conditions required in the Trotter-Kato theorem (see e.g.~Conditions {\bf (A1)} and {\bf (A2)} in Theorem~\ref{ParisVI_thm} below)  have been analyzed  in \cite{Banks_al78} and \cite{banks1979spline}. 

For the case of Legendre polynomials, technical complications have been encountered to check these conditions either in the setting of Galerkin approximation \cite{Kappel86} or in the setting of tau-method \cite{Ito_Teglas86} largely due to the fact that the basis functions do not live in the domain of $\mathcal{A}$. As noted in \cite[p.~168]{Kappel86} or in \cite[Sect.~5]{Ito_Teglas86}, either $X_N\not\subset D(\mathcal{A})$ or $\Pi_N$ is not orthogonal for the polynomial functions considered in \cite{Kappel86} and \cite{Ito_Teglas86}, respectively.

On the other hand, at a given precision, the use of polynomial basis leads typically to ODE approximations with lower dimensions when compared with those built out of step functions or splines \cite{banks1979spline,Ito_Teglas86}.  \qed

\ei
\er

The problems discussed in (iii) of the above remark already encountered in the linear case,  have limited the applications of polynomial basis for the approximation of nonlinear systems of DDEs.  The above discussion leads naturally to the question whether there exists an orthogonal polynomial basis for which standard approximation results for abstract nonlinear systems such as recalled in Theorem \ref{ParisVI_thm} below, could be applied to the case of nonlinear systems of DDEs.

The next section introduces orthogonal polynomials that will allow  us to answer this question by the affirmative, leading to direct and explicit formulas for the rigorous Galerkin approximations of a broad class of nonlinear  systems of DDEs; see Sections~\ref{Subsect_DDE_Galerkin} and \ref{Sec_Galerkin_analytic}.  As explained next, the key is  to seek for polynomials that live in the domain of $\mathcal{A}$, which is achieved here by seeking for polynomials to be orthogonal for the  inner product  \eqref{H_inner} with a point mass.

\section{Orthogonal polynomials for inner products with a point mass} \label{sect:basis}
The inner product given in \eqref{H_inner} is naturally associated with the measure
\be\label{Eq_dx+point-mass}
\nu(\d \theta)=\d \theta +\delta_0,
\ee
where $\delta_0$ denotes the Dirac measure concentrated at $\theta=0$. 

Orthogonal polynomials with respect to the Lebesgue measure $\d \theta$ or measures having  a smooth density with respect to it,  
has a long history \cite{Szego75}.   The study of orthogonal polynomials with respect to a measure including  a point mass such as given by \eqref{Eq_dx+point-mass} has been studied only lately \cite[Chap.~2.9]{Ismail05}. It was in particular noticed that orthogonal polynomials with respect to such  a measure  can be expressed in terms of  polynomials orthogonal with respect to the smooth part of the measure; see \cite{Uvarov69} for an early contribution on the topic. 

Koornwinder in \cite{Koo84} dealt with the case of orthogonal polynomials on $[-1,1]$ associated with measures given by 
\be\label{Eq_Koorn_dx}
\nu(\d x)= \frac{\Gamma(\alpha+\beta+2)}{2^{\alpha+\beta+1}\Gamma(\alpha+1)\Gamma(\beta+1)}(1-x)^\alpha (1+x)^\beta \d x  + M \delta_{-1}+ N\delta_{1}, \; \alpha, \beta > -1,
\ee 
i.e.~associated with measures having a Jacobi weight on $[-1,1]$ with two point-masses added to the extremities of the interval.

Although many properties --- such as three-term recurrence relationships or differential equations satisfied by such polynomials --- remain valid 
in the case of a measure with a point mass, subtle but important qualitative and quantitative differences arise. For instance, \cite[Thm.~ 3 c)]{Alfaro_al97} shows that the zeroes closest to $1$ of polynomials orthogonal with respect to the measure $\nu$ given in \eqref{Eq_Koorn_dx} converge to $1$ faster than those associated with the standard Jacobi polynomials.

%%%%%
It is our goal to show that orthogonal polynomials with respect to the measure $\nu$ in \eqref{Eq_Koorn_dx} allows us to work within a simple and more direct framework than those found in the literature, for building Galerkin  approximations of DDEs.  Indeed, the approximation of DDEs  by systems of ODEs built  from orthogonal polynomials were not, so far, relying on classical Galerkin schemes as noted in \cite[p.~168]{Kappel86} or in \cite[Sect.~5]{Ito_Teglas86}.

%%%%%%%%%%%%%%%%%%%%%%%%%%

%%%%%%%%%%%%%%%%%%%%%%%%%%
The results given  below correspond to the case $\alpha = \beta = M=0$, and $N = 1$ considered in \cite{Koo84}.

\subsection{Koornwinder polynomials}
We recall next from \cite[Eq.~(2.1)]{Koo84} the following sequence of Koornwinder polynomials $\{K_n\}$ that can be built from 
the Legendre polynomials $L_n$ according to
\be \label{eq:Pn}
K_n(s) := -(1+s)\frac{\d}{\d s} L_n(s) +( n^2 + n + 1) L_n(s), \; s \in [-1, 1], \; n \in \mathbb{N}. 
\ee

As recalled above, this polynomial sequence is known to be orthogonal when a Dirac point-mass at the right endpoint, $\delta_{1}$, is adjoined 
to the Lebesgue measure \cite{Koo84}, in other words
\bea
\int_{-1}^{1} K_n(s) K_m(s) \d \mu (s)& = \frac{1}{2} \int_{-1}^{1} K_n(s) K_m(s) \d x + K_n(1) K_m(1)\\
&=0, \, \mbox{ if } m\neq n.
\eea
This orthogonality property and the main properties satisfied by $\{K_n\}$ on which we will rely on, are 
summarized from \cite{Koo84} in the proposition below.

\bprop \label{thm:Pn}

The polynomial $K_n$ defined in \eqref{eq:Pn} {\mk is of degree $n$} and  admits the following expansion in terms of the Legendre polynomials:
\be \label{eq:Pn2}
K_n(s) = - \sum_{j = 0}^{n-1} (2j+1)L_j(s) + (n^2 + 1) L_n(s), \qquad n \in \mathbb{N};
\ee
and the following normalization property holds:
\be \label{eq:Pn_normalization}
K_n(1) = 1, \qquad n \in \mathbb{N}.
\ee

Moreover, the sequence given by
\be \label{eq:Pn_prod}
\{\mathcal{K}_n := (K_n, K_n(1)) : n \in \mathbb{N}\}
\ee 
forms an orthogonal basis of the  product space 
\be \label{eq:E}
\mathcal{E} := L^2([-1,1); \mathbb{R}) \times  \mathbb{R},
\ee 
where $\mathcal{E}$ is {\mk endowed} with the following inner product:
\be \label{eq:inner_E}
\langle (f, a), (g, b) \rangle_{\mathcal{E}}  = \frac{1}{2} \int_{-1}^1 f(s)g(s) \d s  + ab, \quad (f,a), (g, b) \in \mathcal{E}.
\ee

{\mk Moreover $\Big\{\frac{\mathcal{K}_n}{\|\mathcal{K}_n\|_{\mathcal{E}}}\Big\}$ forms a Hilbert basis of $\mathcal{E}$ where 
the norm $\|\mathcal{K}_n\|_{\mathcal{E}}$ of $\mathcal{K}_n$ induced by  $\langle \cdot, \cdot \rangle_{\mathcal{E}}$  possesses the following analytic expression:}
\be \label{eq:Pn_norm}
\|\mathcal{K}_n\|_{\mathcal{E}} = \sqrt{\frac{(n^2+1)((n+1)^2+1)}{2n+1}}, \qquad n \in \mathbb{N}.
\ee

\eprop

\bp
Based on \eqref{eq:Pn}, the proof consists essentially of algebraic manipulations relying on the following standard properties of the Legendre polynomials \cite[Sect.~3.3]{Shen_al11}:
\bi
\item Orthogonality:
\be \label{eq:Ln_orth}
\int_{-1}^1 L_m(s) L_n(s) \d x  = \frac{2}{2n+1} \delta_{mn}, \qquad m,n \in \mathbb{N},
\ee 
where $\delta_{mn}$ denotes the Kronecker delta. 

\medskip

\item Normalization:
\be \label{eq:Ln_normalization}
L_n(1) = 1, \qquad n \in \mathbb{N}.
\ee

\item Three-term recurrence relation: 
\be \label{eq:Ln_recur}
 (n+1) L_{n+1}(s)  = (2n+1) s L_n(s) - n L_{n-1}(s), \; s\in [-1,1], \; n \ge 1,
\ee
{\mk where} the first two Legendre polynomials are  {\mk given by}
\be
L_0 \equiv 1 \qquad \text{and} \qquad  L_1(s) = s.
\ee

\item First order derivative recurrence relation:

\be \label{eq:dLn}
\frac{\d L_n}{\d s}(s)  = \sum_{k \in I_n} (2k+1)L_k(s), 
\ee
where 
\be \label{eq:idx_n}
I_n:=\{k \in \mathbb{N} : 0\le k \le n-1, k+n \text{ is odd}\}.
\ee

\ei
Standard density arguments, outlined in Appendix~\ref{sect:thm_Pn_proof}, allow us then to conclude the proof.
\ep

\subsection{Rescaled Koornwinder basis} \label{Sect_rescaled_basis}
From the original Koornwinder basis given on the interval $[-1, 1]$, orthogonal polynomials on the interval $[-\tau, 0]$ for the inner product \eqref{H_inner} can now be easily obtained by using a simple linear transformation $\mathcal{T}$ defined by:
\be \label{eq:linear_transf}
\mathcal{T} \colon [-\tau, 0] \rightarrow [-1, 1], \qquad \theta \mapsto 1 + \frac{2 \theta }{\tau}. 
\ee
Indeed, for $K_n$ given in \eqref{eq:Pn2}, let us define the polynomial $K_n^\tau$ by 
\bea \label{eq:Pn_tilde}
K^\tau_n\colon  [-\tau, 0] & \rightarrow \mathbb{R}, \\
\theta & \mapsto  K_n \Bigl( 1 + \frac{2 \theta }{\tau} \Bigr), \qquad n \in \mathbb{N}.
\eea
Since the sequence $\{\mathcal{K}_n = (K_n, K_n(1)) : n \in \mathbb{N}\}$ forms an orthogonal basis for $\mathcal{E}$ (cf.~Proposition~\ref{thm:Pn}), it follows then that the polynomial sequence 
\be \label{eq:Pn_tilde_prod}
\{\mathcal{K}_n^\tau := (K_n^\tau, K_n^\tau(0)) : n \in \mathbb{N}\}
\ee
forms an orthogonal basis for the space $\mathcal{H} = L^2([-\tau,0); \mathbb{R}) \times  \mathbb{R}$ endowed with the inner product $\langle \cdot, \cdot \rangle_{\mathcal{H}}$ given in \eqref{H_inner} for $d=1$.

Since $K_n(1)=1$ from \eqref{eq:Pn_normalization}, we have 
\be\label{Eq_normalization}
K_n^\tau(0)  = 1.
\ee
Moreover, by applying the transformation $\mathcal{T}$, we get trivially that
\be\label{Eq_inv_innerproduct} 
\|\mathcal{K}_n^\tau \|_{\mathcal{H}}  = \| \mathcal{K}_n \|_{\mathcal{E}}.
\ee
We have then the following fundamental lemma.

\bl\label{Fundamental_lemma}

The rescaled Koornwinder polynomials $\{K_j^{\tau}\}_{j\geq 0}$  satisfy the following properties:
\be \label{Eq_identity1_0}
\boxed{
\sum_{j=0}^{\infty} \frac{K^\tau_j}{\|\mathcal{K}_j^\tau \|_{\mathcal{H}}^2}=0,  \quad \text{ in the $L^2$ sense},}
\ee
and
\be \label{Eq_identity2_0}
\boxed{
\sum_{j=0}^{\infty} \frac{1}{\|\mathcal{K}_j^\tau \|_{\mathcal{H}}^2}=1.}
\ee

Moreover, each function in $L^2([-\tau, 0]; \mathbb{R})$ enjoys the following decomposition in terms of the Koornwinder polynomials $K_j^\tau$:
\be \label{L2_decomp_0}
\boxed{
f  = \sum_{j=0}^\infty \frac{\langle f, K_j^\tau \rangle_{L^2} }{\tau \|\mathcal{K}_j^\tau \|_{\mathcal{H}}^2 }  K_j^\tau,\qquad \Forall f \in L^2([-\tau, 0]; \mathbb{R}).}
\ee

\el

\bp
For any $\Psi=(\Psi^D,\Psi^S) \in \mathcal{H}$, we have\footnote{Note that the equality in \eqref{Eq_decomp} holds in the sense that $ \Big \|\Psi - \sum_{j=0}^{\infty} \frac{\langle \Psi,\mathcal{K}^\tau_j \rangle_{\mathcal{H}}}{\|\mathcal{K}^\tau_j\|_{\mathcal{H}}^2}\mathcal{K}_j^\tau \Big\|_\mathcal{H}$ = 0, which is equivalent to $\Big \|\Psi^D - \sum_{j=0}^{\infty} \frac{\langle \Psi,\mathcal{K}^\tau_j \rangle_{\mathcal{H}}}{\|\mathcal{K}^\tau_j\|_{\mathcal{H}}^2} K_j^\tau \Big\|_{L^2} = 0$ and 
$\Big |\Psi^S - \sum_{j=0}^{\infty} \frac{\langle \Psi,\mathcal{K}^\tau_j \rangle_{\mathcal{H}}}{\|\mathcal{K}^\tau_j\|_{\mathcal{H}}^2}\Big| = 0$.}
\bea\label{Eq_decomp}
\Psi &=\sum_{j=0}^{\infty} \frac{\langle \Psi,\mathcal{K}^\tau_j \rangle_{\mathcal{H}}}{\|\mathcal{K}^\tau_j\|_{\mathcal{H}}^2}\mathcal{K}_j^\tau\\
&  =\sum_{j=0}^{\infty} \Big(\frac{1}{\tau}\langle \Psi^D,K_j^\tau \rangle_{L^2}+\Psi^S K_j^\tau(0)\Big) \frac{\mathcal{K}_j^\tau }{\|\mathcal{K}_j^\tau \|_{\mathcal{H}}^2}.
\eea

Now, let $\Psi^D$ to be  the zero-function on $[-\tau,0]$ and $\Psi^S$ to be 1. 
For such a $\Psi$, by equalizing respectively the $D$-components and $S$-components  of the RHS and LHS of \eqref{Eq_decomp}, one then obtains from \eqref{Eq_normalization} that
\be
\sum_{j=0}^{\infty}  \frac{K^\tau_j}{\|\mathcal{K}_j^\tau \|_{\mathcal{H}}^2}=0, \quad \text{ in the $L^2$ sense},
\ee
and
\be
\sum_{j=0}^{\infty}  \frac{1}{\|\mathcal{K}_j^\tau \|_{\mathcal{H}}^2}=1.
\ee

The decomposition of $L^2$ functions given in \eqref{L2_decomp_0} follows directly from \eqref{Eq_decomp} by considering $\Psi := (f, 0) \in \mathcal{H}$. Again, the equality holds in the $L^2$ sense.

\ep

\begin{figure}[hbtp]
   \centering
\includegraphics[height=0.4\textwidth,width=.75\textwidth]{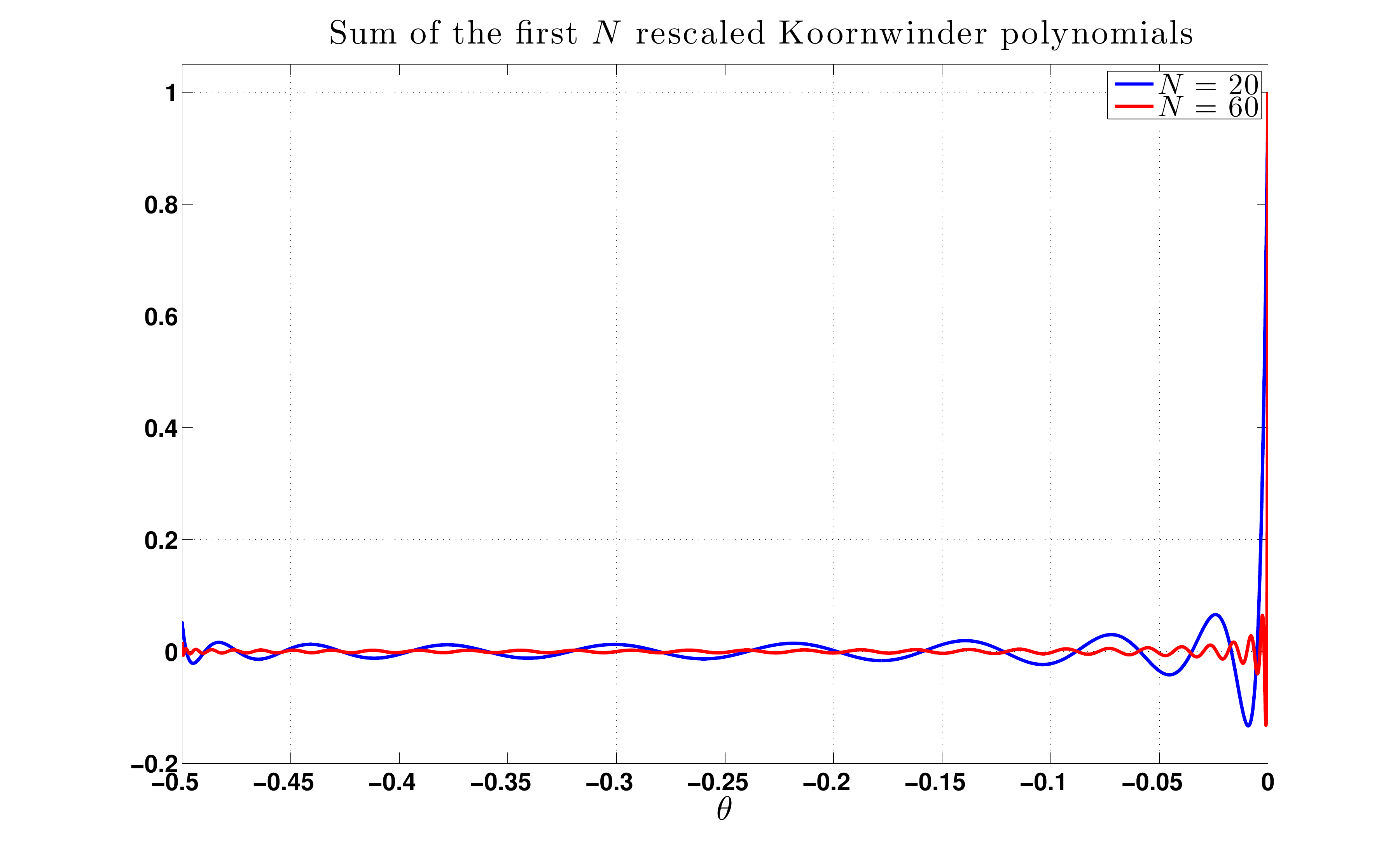}
  \caption{{\footnotesize Sum of the first $N$ rescaled Koornwinder polynomials:  blue curve corresponds to $N=20$, and red curve to $N=60$.}}   \label{fig:Cancel_prop}
\end{figure}

 As an illustration of the identities \eqref{Eq_identity1_0} and \eqref{Eq_identity2_0}, Figure~\ref{fig:Cancel_prop} displays numerical computations of the partial sum $\sum_{j=0}^{N-1} \frac{K^\tau_j}{\|\mathcal{K}_j^\tau \|_{\mathcal{H}}^2}$ for $N=20$ and $N=60$, in the case $\tau = 0.5$.

\needspace{1\baselineskip}
\br \label{Rmk_KoornwinderBasis}
\hspace*{2em}  \vspace*{-0.4em}
\bi

\item[(i)] Note that the continuity condition, $\lim_{\theta \rightarrow 0^-} \Psi^D(\theta) = \Psi^S $,  required in \eqref{D_of_A} in order that $\Psi \in D(\mathcal{A})$, is here naturally satisfied by the Koornwinder basis function $\mathcal{K}_n^\tau = (K_n^\tau, K_n^\tau(0))$. It constitutes thus, for the inner product \eqref{H_inner}, an orthogonal polynomial basis whose elements live in the domain of the linear operator $\mathcal{A}$ given in \eqref{Def_A}. 

As explained at the beginning of Section~\ref{sect:basis}, the key element for such a construction relies on the inclusion of a point mass adjoined to the continuous part of the measure. When this point mass is absent, the corresponding orthogonal polynomials are (rescaled) Legendre polynomials. The associated basis in this latter case is given by (cf.~\cite{Kappel86}) 
\be \label{Legendre-based basis}
\mathfrak{B} := \{ \psi_1:= (0_{\mathcal{H}}, 1) \} \cup \{\psi_n:= (L^\tau_{n-2}, 0)  \ \vert \ n=2, 3, \cdots \},
\ee
where $0_\mathcal{H}$ denotes the zero function on $\mathcal{H}$, and $L^\tau_n$ is the Legendre polynomial of degree $n$ defined on the interval $[-\tau, 0]$. Clearly, none of the elements in $\mathfrak{B}$ belongs to $D(\mathcal{A})$ since $\lim_{\theta \rightarrow 0^-} L^\tau_n(\theta) \neq 0$.

\item[(ii)] The fact that the Koornwinder basis functions live in $D(\mathcal{A})$ allows us to construct {\it standard} Galerkin approximations; whereas, extra correction terms are required in the Galerkin approximation built from the Legendre-based basis given in \eqref{Legendre-based basis} (see e.g.~\cite[p.~169]{Kappel86}). Moreover, technical complications such as pointed out in Remark~\ref{Rmk_problems_to_overcome} iii) do not take place for the case of Koornwinder basis. Indeed, it will be shown in Section~\ref{Subsect_DDE_Galerkin} that the properties of the Koornwinder polynomials such as summarized in Corollary~\ref{Fundamental_lemma} as well as the vectorized version given by Corollary~\ref{Super_Fundamental_lemma} below, turn out to be sufficient to obtain finite-time uniform approximation results of the semigroup generated by the linear operator $\mathcal{A}$; see Lemmas~\ref{Lem_A2} and \ref{Lem_A1}. 

\item[(iii)] It is also worth mentioning that Corollary~\ref{Fundamental_lemma} and   Corollary~\ref{Super_Fundamental_lemma} as well as the rigorous convergence results presented in Section~\ref{Subsect_DDE_Galerkin} are not limited to the case of Koornwinder basis constructed here. Given any 
polynomial basis on $[-\tau, 0]$ that are orthogonal with respect to a measure of the form $\nu(\mathrm{d} \theta) = \mathrm{d} \rho(\theta) + \delta_0$ with $\rho$ being a positive non-decreasing function on $[-\tau, 0]$, the aforementioned results would still hold. We refer to \cite{Uvarov69} for the construction of such polynomials when orthogonal polynomials with respect to $\widetilde{\nu}(\mathrm{d} \theta) = \mathrm{d} \rho(\theta)$ are known.  \qed

\ei 
\er

%%%%%%%%%%%%%%%%%%%%%%%%%%%%

\subsection{Vectorization of Koornwinder polynomials}\label{Sec_Vectorization}
We introduce here a generalization of the Koornwinder polynomials that will turn out to be useful for 
the approximation of nonlinear DDE systems. 

The purpose is here to build from the Koornwinder polynomials introduced above,  linear subspaces $\mathcal{H}_{N}$ that approximate $\mathcal{H} := L^2([-\tau,0); \mathbb{R}^d) \times  \mathbb{R}^d$, for $d>1$.

Each function $\Psi$ in $\mathcal{H}$ has here $d$-components that can be, each, approximated by a series of Koornwinder polynomials as in \eqref{Eq_decomp}. If we restrict such an approximation to the first $N$ Koornwinder polynomials, $\mathcal{H}_{N}$ becomes then an $N\times d$-dimensional subspace; see \eqref{subspace_HNd} below. 

Our goal is also here to introduce a vectorization  of Koornwinder polynomials which allows for a natural extension of Lemma \ref{Fundamental_lemma} in the case $d>1$. This extension of Lemma \ref{Fundamental_lemma} will be particularly useful to provide finite-dimensional approximation of the linear part of systems of DDEs; see Lemma \ref{Lem_A2}. 

To do so, given $j\in\{1,\cdots,Nd\}$,  we can associate a Koornwinder polynomial  of degree $j_q \in\{0,\cdots,N-1\}$, as follows
\be\label{index_relation}
j=d j_q +j_r,
\ee
where $j_r \in \{1,\cdots,d\}$ is given by
\be\label{index_relationb}
j_r := \begin{cases}
\mathrm{mod}(j, d), & \text{if } \mathrm{mod}(j, d) \neq 0, \\
d,  & \text{otherwise.} 
\end{cases} 
\ee

Let us introduce now the following $d$-dimensional mapping from $[-\tau,0]$ to $\mathbb{R}^d$
\be \label{vectorized_K}
\mathbf{K}^\tau_{j}(\theta):= (\underbrace{0, \cdots, 0}_{\text{$j_r-1$ entries}}, K^\tau_{j_q}(\theta), \underbrace{0, \cdots, 0}_{\text{$d-j_r$ entries}})^\tr, \qquad  \theta \in [-\tau,0].
\ee

The vector $\mathbf{K}^\tau_{j}(\theta)$ is then nothing else than a $d$-dimensional  canonical vector whose  $j_r^{th}$-entry is given by the value at $\theta$ of the (rescaled) Koornwinder polynomial of degree $j_q$; the integers $j_q$ and $j_r$ being related to $j$ according to \eqref{index_relation}-\eqref{index_relationb}. Based on these vectorized (rescaled) Koornwinder polynomials $\mathbf{K}^\tau_{j}$, we also introduce 
\be\label{Eq_superKoor}
\mathbb{K}^\tau_{j}:= \big( \mathbf{K}^\tau_{j}, \mathbf{K}^\tau_{j}(0) \big), \qquad  j \ge 1. 
\ee

In the remaining part of this section, we summarize some key properties of $\mathbf{K}^\tau_{j}$ and $\mathbb{K}^\tau_{j}$ for later usage. Hereafter, we use $\mathcal{H}_1$ to denote the space $\mathcal{H}$ defined in \eqref{H_space} for the case $d=1$, i.e.,
\be
\mathcal{H}_1 = L^2([-\tau,0); \mathbb{R}) \times  \mathbb{R},
\ee
which is again endowed with the inner product given in \eqref{H_inner} (still with $d=1$).

Since the sequence $\{\mathcal{K}^\tau_j = (K_j^\tau, K_j^\tau(0)) : j \in \mathbb{N}\}$ forms an orthogonal basis for $\mathcal{H}_1$ (cf.~Section~\ref{Sect_rescaled_basis}), one can readily check that $\{\mathbb{K}^\tau_j : j \in \mathbb{N}^*\}$
forms an orthogonal basis for the space $\mathcal{H}$. 

Note also that 
\be\label{Eq_norm_superKoor}
\|\mathbb{K}^\tau_{j}\|_\mathcal{H} = \|\mathcal{K}^\tau_{j_q}\|_{\mathcal{H}_1}, \qquad j \in \mathbb{N}^*.
\ee

Given this vectorization of Koornwinder polynomials, we can now formulate the following extension of Lemma \ref{Fundamental_lemma} that summarizes the key properties of the $\mathbf{K}_j$'s which will be used for the rigorous approximation of nonlinear systems of DDEs such as described in Section \ref{Subsect_DDE_Galerkin}.

\bl\label{Super_Fundamental_lemma}

The vectorized rescaled Koornwinder  polynomials  $\{\mathbf{K}^\tau_j \}_{j\geq 1}$ satisfy the following properties:
\be \label{Eq_identity1}
\boxed{
\sum_{j=1}^{\infty} \frac{   \langle \alpha, \mathbf{K}^\tau_j(0) \rangle }{\|\mathcal{K}_{j_q}^\tau \|_{\mathcal{H}_1}^2} \mathbf{K}^\tau_j =0  \quad \text{ in the $L^2([-\tau,0); \mathbb{R}^d)$ sense},  \quad  \Forall \alpha \in \mathbb{R}^d,} 
\ee
and
\be \label{Eq_identity2}
\boxed{
\sum_{j=1}^{\infty} \frac{ \langle \alpha, \mathbf{K}^\tau_j(0) \rangle }{\|\mathcal{K}_{j_q}^\tau \|_{\mathcal{H}_1}^2}  \mathbf{K}^\tau_j(0)= \alpha, \quad  \Forall \alpha \in \mathbb{R}^d.}
\ee

Moreover, each function in $L^2([-\tau, 0]; \mathbb{R}^d)$ enjoys the following decomposition in terms of the vectorized Koornwinder polynomials $\mathbf{K}_j^\tau$:
\be \label{L2_decomp}
\boxed{
f  = \sum_{j=1}^\infty \frac{\langle f,\mathbf{K}_j^\tau \rangle_{L^2} }{\tau \|\mathcal{K}_{j_q}^\tau \|_{\mathcal{H}_1}^2 }  \mathbf{K}_j^\tau,\qquad \Forall f \in L^2([-\tau, 0]; \mathbb{R}^d);}
\ee
and the following identity holds:
\be \label{Eq_identity3}
\boxed{
\sum_{j=1}^\infty \frac{\langle f,\mathbf{K}_j^\tau \rangle_{L^2} }{\tau \|\mathcal{K}_{j_q}^\tau \|_{\mathcal{H}_1}^2 }  \mathbf{K}_j^\tau(0) = 0,\qquad \Forall f \in L^2([-\tau, 0]; \mathbb{R}^d).}
\ee

\el

\bp

 The above identities can be obtained by using the same type of reasoning as given in the proof of Lemma~\ref{Fundamental_lemma} for the scalar case. 

Indeed, by noting that $\{\mathbb{K}^\tau_j  \,:\,  j \in \mathbb{N}^*\}$ forms an orthogonal basis of $\mathcal{H}$, any $\Psi \in \mathcal{H}$ admits the following decomposition:
\bea\label{Eq_decomp_vec}
\Psi &=\sum_{j=1}^{\infty} \frac{\langle \Psi,\mathbb{K}^\tau_j \rangle_{\mathcal{H}}}{\|\mathbb{K}^\tau_j\|_{\mathcal{H}}^2}\mathbb{K}_j^\tau\\
&  =\sum_{j=1}^{\infty} \Big(\frac{1}{\tau}\langle \Psi^D,\mathbf{K}_j^\tau \rangle_{L^2}+ \langle \Psi^S,  \mathbf{K}_j^\tau(0) \rangle \Big) \frac{\mathbb{K}_j^\tau}{\|\mathcal{K}_{j_q}^\tau \|_{\mathcal{H}_1}^2},
\eea
where we have used the identity \eqref{Eq_norm_superKoor} in the last equality above. 

Now, let $\Psi^D \in L^2([-\tau, 0]; \mathbb{R}^d)$ to be  the zero-function and $\Psi^S$ to be an arbitrary vector $\alpha \in \mathbb{R}^d$. For such a $\Psi$, by equalizing respectively the $D$-components and $S$-components  of the RHS and LHS of \eqref{Eq_decomp_vec}, we obtain respectively \eqref{Eq_identity1} and \eqref{Eq_identity2}.

The identities \eqref{L2_decomp} and \eqref{Eq_identity3}
 also follow directly from \eqref{Eq_decomp_vec} by considering $\Psi^D = f$ and $\Psi^S = 0 \in \mathbb{R}^d$.

\ep

\section{Galerkin approximation: Rigorous results}\label{Sec_Galerkin_approx}

In this section, we establish the convergence of the Galerkin scheme based on the 
rescaled and vectorized Koornwinder polynomials  of Section~\ref{Sec_Vectorization}. These convergence results apply, as we shall see, to a broad class of nonlinear systems of DDEs.

As mentioned in the Introduction and in Remark \ref{Rmk_problems_to_overcome}-(iii), the advantage of the Koornwinder basis relies on the facts that the constitutive basis functions are orthogonal and belong each to the domain of the linear operator associated with a given DDE. In particular,  there is no discontinuity at the right end point for each basis element, by construction; see Section \ref{sect:basis}. Thanks to these properties of the basis functions, convergence results for the associated Galerkin systems can be derived in a straightforward fashion (see Corollary \ref{Cor_DDE_global_Lip}) and under useful criteria on the nonlinear terms (see Corollaries \ref{Cor_DDE_local_Lip_Case1} and \ref{Cor_DDE_local_Lip_Case2}),  compared to other Galerkin schemes built from other bases; see, e.g., \cite{Kappel86,Kappel_al87, Vyasarayani12,Wahi_al05} and references therein.  

In the following, we first present in Section \ref{Subsect_ODE_Galerkin} a general convergence result for Galerkin approximations of  abstract nonlinear ODEs in Hilbert spaces by relying essentially on the theory of semigroups and the Trotter-Kato theorem \cite[Thm.~4.5, p.~88]{Pazy83}.  The result is then applied to the DDE context in Section \ref{Subsect_DDE_Galerkin}. 
General examples are provided in Section \ref{Sec_examples}.

\subsection{Galerkin approximations of nonlinear ODEs in Hilbert spaces} \label{Subsect_ODE_Galerkin}
We first present a general convergence result for Galerkin approximations of abstract nonlinear differential equations in a Hilbert space $X$, endowed with a norm $\|\cdot\|_X$.  The mathematical setting is somewhat classical but we recall it below for the reader's convenience and later use. 

In that respect, we assume in this Section the linear operator $\mathcal{L}$ to be the  infinitesimal  generator of a
$C_0$-semigroup of bounded linear operators $T(t)$ on $X$. Recall that in that case the domain $D(\mathcal{L})$ of $\mathcal{L}$ is dense in $X$ and that $\mathcal{L}$ is a closed operator; see  \cite[Cor.~2.5, p.~5]{Pazy83}. 

Under these assumptions, recall that there exists $M\geq 1$ and $\omega \geq 0$  \cite[Thm.~2.2, p.~4]{Pazy83} such that
\be\label{Eq_control_T_t}
\|T(t)\| \le M e^{\omega t},  \qquad t \ge 0,
\ee
where $\|\cdot \|$ denotes the operator norm subordinated to $\|\cdot\|_X$.

We are concerned with finite-dimensional approximations of the following initial-value problem:
\bea \label{ODE}
\frac{\d u}{\d t} &= \mathcal{L} u + \mathcal{G}(u), \\
u(0) &= u_0,
\eea
where $u_0 \in X$.

A {\it mild solution} of \eqref{ODE} over $[0,T]$, will be any function $u\in C([0,T],X)$ such that for $u_0\in X$,
\be\label{Eq_mild}
u(t)=T(t)u_0 + \int_0^t T(t-s) \mathcal{G}(u(s)) \d s.
\ee

Let $\{X_N: N \in \mathbb{N}\}$ be a sequence of subspaces of $X$ associated with {\it orthogonal  projectors}
\be
\Pi_N: X \rightarrow X_N,
\ee
such that 
\be\label{Eq_identity_approx}
\|\Pi_N-\mbox{Id}_X\|\underset{N\rightarrow \infty}\longrightarrow 0,
\ee
and
\be\label{Eq_XN_in_domain}
X_N\subset D(\mathcal{L}), \; \forall \, N.
\ee

The corresponding Galerkin approximation of \eqref{ODE} associated with $X_N$ is then given by:
\bea \label{ODE_Galerkin}
\frac{\d u_N}{\d t} &= \mathcal{L}_N u_N + \Pi_N \mathcal{G}(u_N), \\
u_N(0) &= \Pi_N u_0, \; u_0\in X,
\eea
where  
\be\label{Def_LN}
\mathcal{L}_N := \Pi_N \mathcal{L} \Pi_N : X \rightarrow X_N. 
\ee
In particular, the domain 
$D(\mathcal{L}_N)$ of $\mathcal{L}_N$ is $X$, because of \eqref{Eq_XN_in_domain}. 

As we will see in Section \ref{Subsect_DDE_Galerkin}, the choice of vectorized Koornwinder polynomials as a basis function will allow us 
to define  subspaces $X_N$ naturally associated with orthogonal projectors $\Pi_N$ that satisfy the above properties in contrast to other polynomial functions used for (non-standard) Galerkin approximation or other Legendre-tau approximations  of systems of DDEs used so far; see e.g.~\cite{Ito_Teglas86,Kappel86}. See also Remark~\ref{Rmk_problems_to_overcome}-iii).  

These nice properties will allow us also to rely on the following general convergence result regarding 
{\it standard} Galerkin schemes, for the case of nonlinear systems of DDEs; see Section \ref{Subsect_DDE_Galerkin} below.

\vspace{1ex}
\bt \label{ParisVI_thm}
Let $\mathcal{L}$ and $\{X_N\}_{N\geq 0}$ be as described above. 
Assume furthermore the following set of assumptions:
\bi

\item[{\bf (A1)}]  For each $N\in \mathbb{N}$, the linear flow $e^{\mathcal{L}_N t}:X_N \rightarrow X_N$ extends to a $C_0$-semigroup $T_N(t)$ on $X$. Furthermore the following uniform bound is satisfied by the family $\{T_N(t)\}_{N\geq 0, t\geq0}$
\be \label{Eq_control_linearflow}
\quad \|T_N(t)\| \le M e^{\omega t}, \quad N\geq 0, \; \quad t \ge 0,
\ee
where $\|T_N(t)\|=\sup\{\|T_N(t)x\|_X, \; \|x\|_X\leq 1, x\in X\}$.

\item[{\bf (A2)}] 
The following convergence holds
\be \label{Eq_L_Approx}
\lim_{N \rightarrow \infty} \|\mathcal{L}_N  \phi - \mathcal{L} \phi \|_X = 0, \quad \Forall \phi \in D(\mathcal{L}).
\ee

\item[{\bf (A3)}] $\mathcal{G}$ is globally Lipschitz.

\ei
Then for any $u_0 \in X$, there exists a unique mild solution of  \eqref{ODE} and such a solution can be approximated uniformly 
on each bounded interval $[0, T]$ by the sequence $\{t\mapsto u_N(t; \Pi_N u_0)\}_{N\geq 0}$ of  mild solutions obtained from \eqref{ODE_Galerkin}, i.e.:
\be \label{uniform_conv_est_ODE}
\lim_{N\rightarrow \infty} \sup_{t \in [0, T]} \|u_N(t; \Pi_N u_0) - u(t; u_0)\|_X = 0,  \qquad \Forall T > 0.
\ee

\et

\vspace{1ex}
\bp

Recall that the existence and uniqueness of solutions to Eq.~\eqref{Eq_mild} emanating from any initial data $u_0 \in X$ can be proved by a fixed point argument in $C([0,T],X)$ as in the proof of e.g.~\cite[Prop.~4.3.3]{Cazenave_al98}, by relaxing the semigroup of contractions requirement therein to the $C_0$-semigroup setting adopted here; see also 
\cite[Thm.~6.1.1]{Lunardi04}.  

Given $u_0\in X$, let $u$ be thus the unique mild solution of Eq.~\eqref{ODE}.  
By the variation-of-constants formula applied to Eq.~\eqref{ODE_Galerkin} we have on the other hand, for $0\leq t\leq T$,
\bea
u_N(t) = e^{\mathcal{L}_N t} \Pi_N u_0 + \int_0^t e^{\mathcal{L}_N (t -s )} \Pi_N \mathcal{G}(u_N(s)) \d s.
\eea

Then $v_N(t)=u(t)  - u_N(t)$ satisfies 
\bea \label{eq:residual}
v_N(t)&=  T(t) u_0 - e^{\mathcal{L}_N t} \Pi_N u_0 +  \int_0^t T(t-s) \mathcal{G}(u(s)) \d s - 
\int_0^t e^{\mathcal{L}_N (t -s )} \Pi_N \mathcal{G}(u_N(s)) \d s  \\
& =  T(t) u_0 - e^{\mathcal{L}_N t} \Pi_N u_0 +  
\int_0^t \big( T(t-s) -  e^{\mathcal{L}_N (t -s )} \Pi_N  \big)  \mathcal{G}(u(s)) \d s  \\
& \hspace{10em} +  \int_0^t e^{\mathcal{L}_N (t -s )} \Pi_N \big( \mathcal{G}(u(s)) - \mathcal{G}(u_N(s)) \big ) \d s.
\eea
Let us introduce 
\bea
 r_N(s) & := \|u(s)  - u_N(s) \|_X, \\
 \epsilon_N(u_0) & := \sup_{t\in[0, T]} \|T(t) u_0 - e^{\mathcal{L}_N t} \Pi_N u_0\|_X, \\
 d_N(s) & := \sup_{t\in[s, T]} \| \big( T(t-s) -  e^{\mathcal{L}_N (t -s )} \Pi_N \big) \mathcal{G}(u(s)) \|_X.
\eea

We obtain  then  from \eqref{eq:residual} that
\bea
 r_N(t)  & \le \epsilon_N(u_0) + \int_0^t d_N(s) \d s + \int_0^t \|e^{\mathcal{L}_N (t -s )} \Pi_N \big( \mathcal{G}(u(s)) - \mathcal{G}(u_N(s)) \big )\|_X \d s \\
& \le \epsilon_N(u_0) + \int_0^t d_N(s) \d s + M \mathrm{Lip}(\mathcal{G}) \int_0^t e^{\omega (t -s )}  r_N(s) \d s \\
& \le \epsilon_N(u_0) + \int_0^T d_N(s) \d s + M \mathrm{Lip}(\mathcal{G}) e^{\omega T} \int_0^t  r_N(s) \d s,
\eea
where we have used the global Lipschitz condition on $\mathcal{G}$ and \eqref{Eq_control_linearflow} to derive the second inequality.

It follows then from Gronwall's inequality that
\be \label{rN_est}
 r_N(t) \le \Big( \epsilon_N(u_0) + \int_0^T d_N(s) \d s \Big) e^{M \mathrm{Lip}(\mathcal{G}) T e^{\omega T} }, \qquad \Forall t \in [0, T].
\ee

We are thus left  with the estimation  of $\epsilon_N(u_0)$ and $\int_0^T d_N(s) \d s $ as $N \rightarrow \infty$. 
The assumptions {\bf (A1)}--{\bf (A2)} allow us to use a version of the Trotter-Kato theorem \cite[Thm.~4.5, p.88]{Pazy83}\footnote{Recall that because $\mathcal{L}$ is the generator of a $C_0$-semigroup $T(t)$ on $X$, it satisfies 
$\|T(t)\| \leq Me^{\omega t},$ and as a consequence the resolvent set of $\mathcal{L}$ contains the interval $]\omega,\infty[$; see \cite[Thm.~5.3 p.~20]{Pazy83}. In particular, for any $f\in \mathcal{H}$ and any $\lambda >\omega$, the equation $(\lambda I -\mathcal{L}) x=f$
possesses a unique solution $x \in D(\mathcal{L})$, which implies in particular that $(\lambda I -\mathcal{L}) D(\mathcal{L})$ is dense in $X$ as required by the version of the Trotter-Kato theorem used here.  This explains why this density  requirement, consequence of our working assumptions, is omitted in the formulation of Theorem \ref{ParisVI_thm}.} 
which implies 
together with \eqref{Eq_identity_approx} that 
\be 
\lim_{N\rightarrow \infty} e^{\mathcal{L}_N t} \Pi_N \phi  = T(t) \phi, \quad \Forall \phi \in X,
\ee
uniformly for $t$ in bounded intervals. 

It follows that
\be \label{epsN_est}
\lim_{N\rightarrow \infty} \epsilon_N(u_0) = 0, \quad \Forall u_0 \in X,
\ee
and that $d_N$ converges point-wisely to zero on $[0,T]$, i.e.
\be \label{dN_est1}
\lim_{N\rightarrow \infty} d_N(s)  = 0, \quad \Forall s \in [0, T].
\ee

On the other hand, from \eqref{Eq_control_T_t}, \eqref{Eq_control_linearflow}, and {\bf (A3)}, we get
\bea
\| \big( T(t-s) -  e^{\mathcal{L}_N (t -s )} \Pi_N \big) &\mathcal{G}(u(s)) \|_X  \le 2 M e^{\omega (t-s)} \|\mathcal{G}(u(s)) \|_X  \\
& \le 2 M e^{\omega (t-s)} \big(\mathrm{Lip}(\mathcal{G}) \|u(s)\|_X + \|\mathcal{G}(0)\|_X\big),
\eea
which implies 
\be  \label{dN_est2}
 d_N(s) \le 2 M e^{\omega T} \big(\mathrm{Lip}(\mathcal{G}) \|u(s)\|_X+ \|\mathcal{G}(0)\|_X\big), \quad \Forall s \in [0, T].
\ee
Since $u\in C([0,T],X)$, $s\mapsto \| u(s)\|_X$ is integrable on $[0,T]$, and  the Lebesgue's dominated convergence theorem allows us then to conclude from \eqref{dN_est1} and \eqref{dN_est2} that
\be \label{dN_est3}
\lim_{N\rightarrow \infty}  \int_{0} ^T d_N(s) \d s  = 0.
\ee
The desired  uniform convergence estimate \eqref{uniform_conv_est_ODE} is then trivially obtained from \eqref{rN_est}. 

\ep

\subsection{Galerkin approximations of nonlinear systems of DDEs}\label{Subsect_DDE_Galerkin}
In this section, given the Hilbert product space 
\bes
\mathcal{H}:=L^2([-\tau, 0); \mathbb{R}^d)\times \mathbb{R}^d, \;\; d\geq 1,
\ees
endowed with the inner product \eqref{H_inner},  we restrict our attention to the following abstract ODE: 
\bea \label{Eq_abstract_ODE_DDE}
\frac{\d u}{\d t} & = \mathcal{A} u + \mathcal{F}(u), \\
\eea
where $\mathcal{F}$ is a nonlinear operator that will be specified later on, and where 
\textemdash\, given $L_D$, a bounded linear operator from $H^1([-\tau, 0); \mathbb{R}^d)$ to $ \mathbb{R}^d$ and, $L_S$, a bounded linear
operator from $\mathbb{R}^d$ to $\mathbb{R}^d$ \textemdash\,  the linear operator $\mathcal{A}$
is given by
\bea \label{Def_A2}
\lbrack \mathcal{A} \Psi \rbrack (\theta) & := \begin{cases}
{\displaystyle \frac{\d^+ \Psi^D}{\d \theta}}, &  \theta \in[-\tau, 0),  \vspace{0.4em}\\ 
{\displaystyle L_S\Psi^S + L_D\Psi^D}, & \theta = 0,
\end{cases}   
\eea
for any $\Psi = (\Psi^D, \Psi^S)$ that lives  in the domain,  $D(\mathcal{A})$, defined as
\be \label{D_of_A2}
D(\mathcal{A}): = \Big \{\Psi \in \mathcal{H} : \Psi^D \in H^1([-\tau, 0); \mathbb{R}^d), \lim_{\theta \rightarrow 0^-}\Psi^D(\theta) =\Psi^S 
\Big \}.
\ee

Such an abstract setting arises naturally in the reformulation of a broad class of nonlinear systems of DDEs
as an abstract ODE in $\mathcal{H}$; see e.g.~\cite{burns1983linear,curtain1995}. Examples of operators $L_D$ depending explicitly on the delay $\tau$ are given below; see \eqref{Def_LD}.

It is well-known that under these assumptions, the operator $\mathcal{A}$ defines a $C_0$-semigroup on $\mathcal{H}$ \cite[Thm.~2.3]{burns1983linear}, and  in particular $\mathcal{A}$ is dense in $\mathcal{H}$ and  is a closed operator. 

We turn now to the definition of the subspaces $X_N$ and $\Pi_N$ of the previous section. 
For each positive integer $N$, we define the $Nd$-dimensional subspace $\mathcal{H}_{N} \subset \mathcal{H}$ to be spanned by the first $Nd$ vectorized Koornwinder polynomials introduced in \eqref{Eq_superKoor}, namely
\be  \label{subspace_HNd}
\mathcal{H}_{N} = \span \Big \{ \mathbb{K}^\tau_{1}, \cdots,  \mathbb{K}^\tau_{Nd} \Big\}.
\ee
As noted in Section \ref{Sec_Vectorization}, these polynomials are orthogonal for the inner product  with a point mass such as defined in \eqref{H_inner}. 

The subspace $\mathcal{H}_{N}$ is thus naturally associated with an orthogonal projector $\Pi_N$, as required in the previous section.
The approximation property \eqref{Eq_identity_approx} is satisfied  due to the density arguments outlined in Appendix~\ref{sect:thm_Pn_proof}. 

Recall finally that by construction $\mathbb{K}^\tau_{j} \in D(\mathcal{A})$ for any $j\in \mathbb{N}^*$, and therefore 
\be\label{Eq_inclusion}
\mathcal{H}_{N}\subset D(\mathcal{A}).
\ee

The corresponding $N$-dimensional Galerkin approximation of Eq.~\eqref{Eq_abstract_ODE_DDE} reads then:
\be\label{Eq_DDE_Galerkin}
\frac{\d u_N}{\d t} = \mathcal{A}_N u_N + \Pi_N \mathcal{F}(u_N), 
\ee
with 
\be\label{Eq_AN}
\mathcal{A}_N := \Pi_N \mathcal{A} \Pi_N,
\ee
which is therefore well defined on $\mathcal{H}$ because of \eqref{Eq_inclusion}.

We are now in  position to check Conditions {\bf (A1)} and {\bf (A2)} of Theorem \ref{ParisVI_thm}.
To check Condition {\bf (A1)}, we will make usage of the following extension of the linear flow $e^{\mathcal{A}_N t}$:
\be\label{Eq_extension}
T_N(t) u=e^{\mathcal{A}_N t} \Pi_N u +(I-\Pi_N) u, \; u\in \mathcal{H}.
\ee
Such an extension leads naturally to a $C_0$-semigroup on $\mathcal{H}$. 
The stability condition \eqref{Eq_control_linearflow}, will require however
some specifications of the operator $L_D$ that will be made clear later.

Condition {\bf (A2)} can be however checked in the general setting by making an appropriate use of  the properties   
of the vectorized Koornwinder polynomials summarized in Lemma \ref{Super_Fundamental_lemma}. More precisely,

\bl   \label{Lem_A2}
Let $\mathcal{H}_N$ be the subspace defined in \eqref{subspace_HNd}. Then for $\mathcal{A}$ defined in \eqref{Def_A2} and $\mathcal{A}_N$ defined in \eqref{Eq_AN} associated with the orthogonal projector $\Pi_N$ onto $\mathcal{H}_N$, we have
\be
\lim_{N \rightarrow \infty} \|\mathcal{A}_{N}   \Psi - \mathcal{A} \Psi \|_\mathcal{H} = 0, \qquad \Forall \Psi \in D(\mathcal{A}).
\ee
\el

\bp
By construction $\mathbb{K}^\tau_{j} \in D(\mathcal{A})$ for each $j\in \mathbb{N}^*$.  Since  $\big\{ \frac{\mathbb{K}^\tau_{j}}{\|\mathbb{K}^\tau_{j}\|_\mathcal{H}}  : j \in \mathbb{N}^*\big\}$ forms a Hilbert basis of $\mathcal{H}$, it suffices to show that
\be \label{Goal_convergence}
\lim_{N \rightarrow \infty} \|\mathcal{A}_{N}  \Psi - \mathcal{A} \Psi \|_\mathcal{H} = 0, \;\;  \Psi \in \underset{k\geq 1}\bigcup  \mathcal{H}_k.
\ee

 We recall from \eqref{Eq_norm_superKoor} that $\|\mathbb{K}^\tau_{j}\|_{ \mathcal{H}} = \|\mathcal{K}^\tau_{j_q}\|_{\mathcal{H}_1}$ for all $j \in \mathbb{N}$. It follows then that the orthogonal projector $\Pi_{N}$ associated with the subspace $\mathcal{H}_{N}$ takes the following explicit form:
\bea \label{DDE_projector}
\Pi_{N} \Psi & = \sum_{j = 1}^{Nd} \frac{ \big \langle \Psi,  \mathbb{K}^\tau_{j} \big \rangle_{\mathcal{H}}}{\|\mathcal{K}^\tau_{j_q}\|_{\mathcal{H}_1}^2} \mathbb{K}^\tau_{j} \\
&  = \sum_{j = 1}^{Nd} \Big(\frac{1}{\tau}\langle \Psi^D, \mathbf{K}_j^\tau \rangle_{L^2}+ \langle \Psi^S,  \mathbf{K}_j^\tau(0) \rangle \Big) \frac{\mathbb{K}_j^\tau }{\|\mathcal{K}_{j_q}^\tau \|_{\mathcal{H}_1}^2}  \\
& = \begin{pmatrix}
p_N & q_N \\
p'_N & q'_N 
\end{pmatrix} 
 \begin{pmatrix}
\Psi^D \\
\Psi^S
\end{pmatrix},
\eea
where the operators $p_N, p'_N, q_N, q'_N$ are defined as following:
\begin{subequations}
\begin{eqnarray}
\hspace{-3em} p_N: L^2([-\tau, 0]; \mathbb{R}^d) \rightarrow L^2([-\tau, 0]; \mathbb{R}^d), & & 
 \Psi^D  \mapsto  \sum_{j = 1}^{Nd} \frac{\langle \Psi^D, \mathbf{K}_j^\tau \rangle_{L^2}}{\tau \|\mathcal{K}_{j_q}^\tau \|_{\mathcal{H}_1}^2} \mathbf{K}_j^\tau; \label{Def_pN} \\
p'_N: L^2([-\tau, 0]; \mathbb{R}^d)  \rightarrow \mathbb{R}^d, & &
 \Psi^D  \mapsto  \sum_{j = 1}^{Nd} \frac{\langle \Psi^D, \mathbf{K}_j^\tau \rangle_{L^2}}{ \tau \|\mathcal{K}_{j_q}^\tau \|_{\mathcal{H}_1}^2}  \mathbf{K}_j^\tau(0);  \label{Def_pN'} \\
 q_N: \mathbb{R}^d \rightarrow L^2([-\tau, 0]; \mathbb{R}^d), &  & 
 \Psi^S  \mapsto  \sum_{j = 1}^{Nd}  \frac{ \langle \Psi^S,  \mathbf{K}_j^\tau(0) \rangle}{\|\mathcal{K}_{j_q}^\tau \|_{\mathcal{H}_1}^2} \mathbf{K}_j^\tau; \label{Def_qN}\\
q'_N: \mathbb{R}^d  \rightarrow \mathbb{R}^d,  & & 
 \Psi^S  \mapsto  \sum_{j = 1}^{Nd}  \frac{ \langle \Psi^S,  \mathbf{K}_j^\tau(0) \rangle}{\|\mathcal{K}_{j_q}^\tau \|_{\mathcal{H}_1}^2}\mathbf{K}_j^\tau(0). \label{Def_qN'}
\end{eqnarray}
\end{subequations}

In the following, we arbitrarily fix $\Phi \in \mathcal{H}_k$ for some integer $k>0$. Now let us choose $N$ such that $Nd \ge k$, then $\Pi_{Nd} \Phi = \Phi$, and we get for each such $N$ 
\be\label{Eq_abstract_AN}
\mathcal{A}_{N} \Phi  = \Pi_{N} \mathcal{A} \Pi_{N}  \Phi 
 = \Pi_{N} \mathcal{A} \Phi  =   \begin{pmatrix}
 p_N  \frac{\d^+}{\d \theta}\Phi^D   +  q_N (L_S \Phi^S + L_D \Phi^D) \vspace{1em}\\
p'_N \frac{\d^+}{\d \theta}\Phi^D  +  q'_N (L_S \Phi^S + L_D \Phi^D)
\end{pmatrix}. 
\ee
We obtain then 
\be \label{eq:AN_residual}
(\mathcal{A} - \mathcal{A}_{N}) \Phi = \begin{pmatrix}
 (I^D - p_N)  \frac{\d^+}{\d \theta}\Phi^D   -  q_N \big(L_S \Phi^S + L_D \Phi^D\big) \vspace{1em}\\
- p'_N \frac{\d^+}{\d \theta}\Phi^D  + (I^S -  q'_N) \big(L_S \Phi^S + L_D \Phi^D \big)
\end{pmatrix}, \; \text{ if } Nd \ge k,
\ee
where $I^D$ and $I^S$ denote the identity maps on $L^2([-\tau, 0]; \mathbb{R}^d)$ and $\mathbb{R}^d$,  respectively.

We show below that the RHS of \eqref{eq:AN_residual} converges to zero. Let us begin with the term $(I^D - p_N)  \frac{\d^+}{\d \theta}\Phi^D$. Note that by comparing the definition of $p_N$ give by \eqref{Def_pN} and the decomposition of $L^2([-\tau, 0]; \mathbb{R}^d)$ functions given by \eqref{L2_decomp}, we see that for each $f \in L^2([-\tau,0]; \mathbb{R}^d)$, the term $p_N f$ is the partial sum of the first $N$ terms in the corresponding decomposition. It follows then that    
\be \label{Eq_pN_residual}
\lim_{N \rightarrow \infty} \|(I^D - p_N) f\|_{L^2} = 0, \quad \Forall f \in L^2([-\tau,0]; \mathbb{R}^d). 
\ee
Since $\Phi \in \mathcal{H}_k \subset D(\mathcal{A})$, it holds that $\frac{\d^+}{\d \theta}\Phi^D \in L^2([-\tau,0]; \mathbb{R}^d)$. We obtain then from \eqref{Eq_pN_residual} that
\be \label{pN_est}
\lim_{N \rightarrow \infty}\Big\|(I^D - p_N) \Big(\frac{\d^+}{\d \theta}\Phi^D \Big) \Big\|_{L^2} = 0.
\ee

\medskip
We turn now to the estimates for $q_N (L_S \Phi^S + L_D \Phi^D)$. By the definition of $q_N$ in \eqref{Def_qN}, we get
\be
q_N \big(L_S \Phi^S + L_D \Phi^D \big) = \sum_{j = 1}^{Nd}  \frac{ \langle L_S \Phi^S + L_D \Phi^D, \mathbf{K}_j^\tau(0)\rangle}{\|\mathcal{K}_{j_q}^\tau \|_{\mathcal{H}_1}^2} \mathbf{K}_j^\tau. 
\ee
It follows then from the identity \eqref{Eq_identity1} that
\be \label{qN_est}
\lim_{N \rightarrow \infty}\Big\|q_N \big( L_S \Phi^S + L_D \Phi^D \big)\|_{L^2} = 0.
\ee

For the term $p'_N \frac{\d^+}{\d \theta}\Phi^D$, since $\frac{\d^+}{\d \theta}\Phi^D \in L^2([-\tau,0); \mathbb{R}^d)$, it follows from the definition of $p'_N$ given in \eqref{Def_pN'} and the identity \eqref{Eq_identity3}  that
\be \label{pN'_est}
\lim_{N \rightarrow \infty} \Big| p'_N \frac{\d^+}{\d \theta}\Phi^D\Big| = 0,
\ee
where $\vert\cdot\vert$ denotes the Euclidean norm of $\mathbb{R}^d$.

By using the identity \eqref{Eq_identity2}, we also get
\be \label{qN'_est}
\lim_{N \rightarrow \infty} \big| (I^S - q'_N) \big(L_S \Phi^S + L_D \Phi^D \big) \big| = 0.
\ee

Now, by using the estimates \eqref{pN_est}, \eqref{qN_est}, \eqref{pN'_est}, and \eqref{qN'_est}, we get from \eqref{eq:AN_residual} that
\be
\lim_{N \rightarrow \infty} \| (\mathcal{A} - \mathcal{A}_{N}) \Phi \|_\mathcal{H} = 0,
\ee
and \eqref{Goal_convergence} follows.

\ep
\vspace{1ex}

\br\label{PDE_rem}
We explain here how the truncated linear operator $\mathcal{A}_N$ defined in \eqref{Eq_abstract_AN} is related to an interesting class of nonlocal linear PDEs.   For the sake of clarity, we discuss the case $d=1$.
For convenience, let us write $v(\theta)=\Phi^D(\theta)$ and, recalling e.g.~Eq.~\eqref{lin_PDE} in the Introduction, replace $\d^+ v/\d \theta$ by $v_{\theta}$ in  \eqref{Eq_abstract_AN}. 

One then obtains that, when $v\in \mathcal{H}_N$,
\be\label{Eq_abstract_AN2}
\begin{pmatrix}
p_N  v_{\theta}\\
p'_N  v_{\theta}  
\end{pmatrix}=\begin{pmatrix}
v_{\theta}\\
v_{\theta}(0)
\end{pmatrix}-\sum_{n=0}^{N-1} \frac{v_{\theta}(0)}{\|\mathcal{K}^\tau_n\|_{\mathcal{H}}^2}\mathcal{K}^\tau_n.
\ee
Next,  we use the expressions of $\mathcal{A}_N$, $p_N$ and $q_N$  --- given, respectively, in \eqref{Eq_abstract_AN}, \eqref{Def_pN} and \eqref{Def_qN}
--- to note that the $D$-component 
$v$ of any solution $u$ of 
\bes
\frac{\d u} {\d t} = \mathcal{A}_N u,
\ees
which emanates from initial data 
taken\footnote{For such initial data, the solution stays in $\mathcal{H}_N$, by the definition of $\mathcal{A}_N$.} in $\mathcal{H}_N$, satisfies the following {\it nonlocal linear PDE}:  
\be\label{nonlocal_PDE}
\partial_t v =\partial_{\theta} v +b_N(\theta)\Big(L_S v(t,0) + L_D v- \partial_{\theta} v\big\vert_{\theta=0}\Big);
\ee
here 
\be\label{bN-coeff}
\displaystyle b_N(\theta)=\sum_{n=0}^{N-1} \frac{K_n^\tau (\theta)}{\|\mathcal{K}_n^{\tau}\|_{\mathcal{H}}^2},
\ee
and the rescaled Koornwinder polynomials $K_n^\tau$ 
are given by \eqref{eq:Pn_tilde}, since $d=1$.
We see therewith that the Galerkin scheme used herein  introduces a 
nonlocal perturbation term with respect to the $D$-component of $\mathcal{A}$ given in
\eqref{Def_A2}.  This perturbation term results from the difference between the $S$-component of  $\mathcal{A}$  and the derivative at $0$, when applied to functions in $\mathcal{H}_N$.

From Lemma \ref{Lem_A2}  above and Lemmas \ref{Lemma_stability_prep} and \ref{Lem_A1} below,  one can infer by the Trotter-Kato theorem that the effects on the solutions of Eq.~\eqref{nonlocal_PDE} of this 
nonlocal perturbation --- which vanishes as $N\rightarrow \infty$, due to \eqref{Eq_identity1_0} --- do not lead to a degenerate situation and that actually these solutions converge 
over finite intervals  to those of the local PDE, $\partial_t v=\partial_\theta v$. This nice convergence is due to the key properties of the Koornwinder polynomials, as summarized in Lemma \ref{Fundamental_lemma} for $d=1$, and in Lemma \ref{Super_Fundamental_lemma} for the multidimensional case; 
see also Fig.~\ref{fig:Cancel_prop} for the nature of the approximation at $\theta = 0$.

To conclude this remark, we return now to the discussion in the Introduction concerning the approximation of discontinuities that arise, for instance, in the first-order derivative of a DDE's solution, cf.~\cite[Appendix A and references therein]{GZT08}.  

In Table \ref{table},  we report the corresponding  
differences over the interval $[0,2]$ 
between the analytic solution of Eq.~\eqref{lin_case} with $\tau=1$ and history $x(t)\equiv 1$,  
on the one hand, and low-dimensional Galerkin approximations obtained by application of the formulas derived hereafter in Section~\ref{Sec_Galerkin_analytic}, on the other.  

\begin{table}[h]
\caption{Errors in Galerkin approximation of Eq.~\eqref{lin_case} \label{table}}
\centering
\begin{tabular}{ccc}
\toprule\noalign{\smallskip}
$N$      &   Max.~error in Galerkin solution   &   Max.~error in 1$^{\textrm{st}}$-order derivative\\ 
&  &           of Galerkin solution\\
\noalign{\smallskip}\hline\noalign{\smallskip}
4       &   $6.9 \times 10^{-3}$                                    &           $5.9\times 10^{-2}$\\
8       &   $9.3\times 10^{-4}$                                   &              $2.2\times 10^{-2}$\\
16      &   $2.3\times 10^{-4}$                                     &           $1.1\times 10^{-2}$\\
32     &   $9.0\times 10^{-5}$                                     &            $5.4\times 10^{-3}$\\
\noalign{\smallskip} \bottomrule 
\end{tabular}
\end{table}

The second column of this table is obviously consistent with the rigorous convergence result of Corollary \ref{Cor_DDE_global_Lip} proved below. The third column shows that, furthermore, the aforementioned discontinuities present in the derivative of the DDE's solutions are well captured by the proposed methodology as well.
\qed

\er

\medskip

In the following, we restrict the linear operator $\mathcal{A}$ defined in \eqref{Def_A2} to be such that 
$L_S: \mathbb{R}^d \rightarrow \mathbb{R}^d$ is a bounded linear operator, and $L_D$ is defined by 
\bea \label{Def_LD}
L_D : H^1([-\tau,0); \mathbb{R}^d)  & \rightarrow \mathbb{R}^d, \\
\Psi^D & \mapsto B \Psi^D(-\tau) +  \int_{-\tau}^0 C(s) \Psi^D(s) \d s,
\eea
with $B: \mathbb{R}^d \rightarrow \mathbb{R}^d$ being a bounded linear operator\footnote{Note that $\Psi^D(-\tau)$ is well-defined for  $\Psi^D\in H^1([-\tau,0); \mathbb{R}^d)$, since the latter Sobolev space is continuously embedded in the space of continuous functions $C([-\tau,0]; \mathbb{R}^d)$; see \cite[Thm.~8.8]{brezis_book}.}, and $C(\cdot) \in L^2([-\tau, 0); \mathbb{R}^{d\times d})$. 

In the following preparatory lemma, Lemma \ref{Lemma_stability_prep}, we recall by means of basic estimates, that the existence of $\omega>0$ such that $\mathcal{A}-\omega I$ is dissipative in $\mathcal{H}$, can be guaranteed.  This fact will be used to establish a stability condition of type \eqref{Eq_control_linearflow} (with $M=1$) for the semigroups $T(t)$ and $T_N(t)$, generated respectively by $\mathcal{A}$ and its finite-dimensional approximation $\mathcal{A}_N$; see Lemma \ref{Lem_A1}. The proofs of these Lemmas are quite straightforward but are reproduced below for the sake of completeness.

\bl \label{Lemma_stability_prep}

Let  $\mathcal{A}$ be defined such as in \eqref{Def_A2} with $L_D$ such as specified in \eqref{Def_LD} and $L_S: \mathbb{R}^d \rightarrow \mathbb{R}^d$ to be a bounded linear operator.  Then
\be
\langle \mathcal{A} \Psi, \Psi \rangle_{\mathcal{H}} \le \omega \|\Psi\|_{\mathcal{H}}^2, \quad \forall\;  \Psi \in D(\mathcal{A}),
\ee
with\footnote{Throughout this article, we will denote indistinguishably by $|\cdot|$, either the Euclidean norm of a vector in $\mathbb{R}^d$, or its subordinated (operator) norm, in the case of a $d \times d$ matrix. It should be clear from the context which norm is used.}  
\be \label{omega}
\omega = \Big(1 + \frac{1}{2 \tau} + |L_S| +  \frac{\tau}{2} |B|^2  + \frac{\tau}{4} \|C\|^2_{L^2}  \Big).
\ee
\el

\bp

Let $\Psi \in D(\mathcal{A})$, then by the definition of $ \mathcal{A}$ given in \eqref{Def_A2}, we have
\be \label{Stab_est0}
\langle \mathcal{A} \Psi, \Psi \rangle_{\mathcal{H}} = \frac{1}{\tau} \int_{-\tau}^0 \Big\langle \frac{\d^+ \Psi^D}{\d \theta}(\theta), \Psi^D(\theta) \Big\rangle  \d \theta + \langle L_S \Psi^S, \Psi^S \rangle  + \langle L_D \Psi^D, \Psi^S \rangle.
\ee

Note that
\bea \label{Stab_est1}
\frac{1}{\tau} \int_{-\tau}^0 \Big\langle \frac{\d^+ \Psi^D}{\d \theta}(\theta), \Psi^D(\theta) \Big\rangle  \d \theta & = \frac{1}{2 \tau} \int_{-\tau}^{0} 
\d |\Psi^D(\theta)|^2  \\
&  = \frac{1}{2 \tau} \Big(|\Psi^S|^2 - |\Psi^D(-\tau)|^2 \Big).
\eea
Using the definition of $L_D$ given in \eqref{Def_LD}, we obtain 
\bea
\langle L_D \Psi^D, \Psi^S \rangle
& = \Big\langle  B \Psi^D(-\tau) +  \int_{-\tau}^0 C(\theta) \Psi^D(\theta) \d \theta, \Psi^S \Big\rangle  \\
& \le |B| |\Psi^D(-\tau)| | \Psi^S| + \|C\|_{L^2} \|\Psi^D\|_{L^2}|\Psi^S|.
\eea
It follows then from Young's inequality that
\bea \label{Stab_est2}
\langle L_D \Psi^D, \Psi^S \rangle & \le \Big( \frac{1}{2 \tau} |\Psi^D(-\tau)|^2  +  \frac{\tau}{2} |B|^2 | \Psi^S|^2 \Big) \\
& \quad + \Big( \frac{1}{\tau}\|\Psi^D\|_{L^2}^2
+ \frac{\tau}{4} \|C\|^2_{L^2} |\Psi^S|^2 \Big).
\eea

Note also that 
\be \label{Stab_est3}
\langle L_S \Psi^S, \Psi^S \rangle  \le |L_s| |\Psi^S|^2.
\ee

Now, by using \eqref{Stab_est1}, \eqref{Stab_est2}, and \eqref{Stab_est3} in \eqref{Stab_est0}, we get
\bea
\langle \mathcal{A} \Psi, \Psi \rangle_{\mathcal{H}} & \le  \Big( \frac{1}{2 \tau} + |L_S| +  \frac{\tau}{2}|B|^2  + \frac{\tau}{4} \|C\|^2_{L^2}  \Big)  |\Psi^S|^2 + \frac{1}{\tau}\|\Psi^D\|_{L^2}^2 \\
& \le  \Big(1 + \frac{1}{2 \tau} + |L_S| +  \frac{\tau}{2}|B|^2  + \frac{\tau}{4} \|C\|^2_{L^2}  \Big)   \Big(| \Psi^S|^2 + \frac{1}{\tau}\|\Psi^D\|_{L^2}^2 \Big) \\
& =  \Big(1 + \frac{1}{2 \tau} + |L_S| +  \frac{\tau}{2}|B|^2  + \frac{\tau}{4} \|C\|^2_{L^2}  \Big) \|\Psi\|_{\mathcal{H}}^2,
\eea
leading thus to the desired estimate.

\ep

%%%%%%%%%%%%%%%%%%%%%%%%%%%

%%%%%%%%%%%%%%%%%%%%%%%%%%%

\bl \label{Lem_A1}

Let  $\mathcal{A}$ be defined such as in \eqref{Def_A2} with $L_D$ such as specified in \eqref{Def_LD} and $L_S: \mathbb{R}^d \rightarrow \mathbb{R}^d$ to be a bounded linear operator.

Then, the linear semigroups $T(t)$ and $T_N(t)$ generated respectively by $\mathcal{A}$ and $\mathcal{A}_N$  defined in \eqref{Eq_AN},  satisfy 
\be\label{stable_estimates}
\|T(t)\| \le e^{\omega t} \quad \text{ and }  \quad \|T_N(t)\| \le e^{\omega t}, \qquad t \ge 0,
\ee
with $\omega$ given by \eqref{omega}.

\el

\bp

Since $T(t)$ is a $C_0$-semigroup with infinitesimal generator  $\mathcal{A}$, we have that $T(t) u_0 \in D(\mathcal{A})$ for all $u_0 \in D(\mathcal{A})$, and that
\be
\frac{\d}{\d t} T(t)u_0 = \mathcal{A} T(t) u_0,   \qquad \Forall u_0 \in \mathcal{A}, \; t \ge 0;
\ee
cf.~\cite[Thm.~2.4 c) p.5]{Pazy83}.

We obtain thus
\be \label{T_est1}
\frac{\d }{\d t }\| T(t) u_0 \|^2_{\mathcal{H}} = 2 \langle \mathcal{A} T(t) u_0, T(t) u_0 \rangle_{\mathcal{H}} \le 2 \omega \|T(t) u_0 \|^2_{\mathcal{H}}, \qquad \Forall u_0 \in D(\mathcal{A}),
\ee
where we have used Lemma~\ref{Lemma_stability_prep} to obtain the last inequality above with $\omega$ given by \eqref{omega}. 

It follows then from Gronwall's inequality that
\be \label{T_est2}
\| T(t) u_0 \|^2_{\mathcal{H}} \le e^{2\omega t} \|u_0\|^2_{\mathcal{H}}, \qquad \Forall  u_0 \in D(\mathcal{A}).
\ee
Since $D(\mathcal{A})$ is dense in $\mathcal{H}$ and $T(t)$ are bounded operators on $\mathcal{H}$, the estimate \eqref{T_est2} still holds for general initial data in $\mathcal{H}$, leading in turn to 
\be \label{stability_A}
\| T(t)\| \le  e^{\omega t}, \qquad t \ge 0.
\ee

The estimate for $T_N$ is also trivial and proceeds as follows. First note that 
by the definition of $T_N$ given by \eqref{Eq_extension}, we have
\bea 
\|T_N(t) u_0\|^2_\mathcal{H} & = \big\langle e^{\mathcal{A}_N t} \Pi_N u_0 + (I - \Pi_N) u_0, e^{\mathcal{A}_N t} \Pi_N u_0 + (I - \Pi_N) u_0 \big\rangle_{\mathcal{H}} \\
& = \big\langle e^{\mathcal{A}_N t} \Pi_N u_0, e^{\mathcal{A}_N t} \Pi_N u_0  \big\rangle_{\mathcal{H}} + \big\langle (I - \Pi_N) u_0, (I - \Pi_N) u_0  \big\rangle_{\mathcal{H}} \\
& = \| e^{\mathcal{A}_N t} \Pi_N u_0 \|^2_{\mathcal{H}} + \| (I - \Pi_N) u_0\|^2_{\mathcal{H}}.
\eea
It follows that
\bea \label{TN_est1}
\frac{\d }{\d t }\| T_N(t) u_0 \|^2_{\mathcal{H}} &= 
\frac{\d }{\d t }\| e^{\mathcal{A}_N t} \Pi_N u_0 \|^2_{\mathcal{H}} \\
& = 2 \langle \mathcal{A}_N e^{\mathcal{A}_N t} \Pi_N u_0, e^{\mathcal{A}_N t} \Pi_N u_0 \rangle_{\mathcal{H}}, \qquad \Forall u_0 \in \mathcal{H}.
\eea
Note also that
\bea  \label{TN_est2}
\langle \mathcal{A}_N \Psi, \Psi \rangle_{\mathcal{H}} & = \langle  \Pi_N \mathcal{A} \Pi_N \Psi, \Psi \rangle_{\mathcal{H}} \\
& = \langle  \Pi_N \mathcal{A} \Pi_N \Psi, \Pi_N \Psi \rangle_{\mathcal{H}} + \langle  \Pi_N \mathcal{A} \Pi_N \Psi, (I- \Pi_N) \Psi \rangle_{\mathcal{H}}\\
& =  \langle \Pi_N  \mathcal{A} \Pi_N \Psi, \Pi_N  \Psi \rangle_{\mathcal{H}} \\
& =  \langle \mathcal{A} \Pi_N \Psi, \Pi_N  \Psi \rangle_{\mathcal{H}} \\
& \le \omega \|\Pi_N \Psi\|_{\mathcal{H}} \\
& \le  \omega \|\Psi\|_{\mathcal{H}}.
\eea
We obtain then from \eqref{TN_est1} that
\bea \label{TN_est3}
\frac{\d }{\d t }\| T_N(t) u_0 \|^2_{\mathcal{H}} & \le 2 \omega \|e^{\mathcal{A}_N t} \Pi_N u_0\|^2_{\mathcal{H}} \\
& \le 2\omega \big( \|e^{\mathcal{A}_N t} \Pi_N u_0\|^2_{\mathcal{H}} + \| (I - \Pi_N) u_0\|^2_{\mathcal{H}} \big) \\
& = 2\omega \|e^{\mathcal{A}_N t} \Pi_N u_0 +(I - \Pi_N) u_0\|^2_{\mathcal{H}} \\
& = 2\omega \|T_N(t) u_0 \|^2_{\mathcal{H}},  \qquad \Forall u_0 \in \mathcal{H}.
\eea

The desired estimate for $\| T_N(t)\|$ can be derived now from \eqref{TN_est3} by using Gronwall's inequality.

\ep

%%%%%%%%%%%%%%%%%%%%%%%%%%%
\br
Note that the estimate about $T_N$ in \eqref{stable_estimates}  shows in particular that solutions of \eqref{nonlocal_PDE} grow at most exponentially with a rate independent of $N$, and stay uniformly bounded over finite intervals.  \qed
\er

%%%%%%%%%
With these preparatory lemmas, we are now in position to obtain as corollaries of Theorem~\ref{ParisVI_thm}, the convergence results for the Galerkin approximation  \eqref{Eq_DDE_Galerkin} of $d$-dimensional nonlinear  systems of DDEs of the form
\be\label{Eq_nln_sys}
\frac{\d \mathbf{x}}{\d t}=L_S \mathbf{x}(t)+ B \mathbf{x}(t-\tau) +  \int_{t-\tau}^t C(s-t) \mathbf{x}(s) \d s + \mathbf{F}\Big(\mathbf{x}(t),\int_{t-\tau}^t  \mathbf{x}(s) \d s\Big),
\ee
where $\mathbf{F}:\mathbb{R}^d\times \mathbb{R}^d  \rightarrow \mathbb{R}^d$, and $L_S$, $B$, and $C$ are as given in \eqref{Def_LD}.

We first sate the result for the case of global    
Lipschitz nonlinearity, keeping in mind that already for the case of scalar DDEs ($d=1$), chaotic dynamics can take place under such a  simple nonlinear setting; see Section \ref{Sec_nearly-brownian} for a numerical illustration.

\vspace{1ex}
\bc \label{Cor_DDE_global_Lip}
Let  $\mathcal{A}$ be defined such as in \eqref{Def_A2} with $L_D$ such as specified in \eqref{Def_LD} and $L_S: \mathbb{R}^d \rightarrow \mathbb{R}^d$ to be a bounded linear operator. 
Assume that the nonlinearity $\mathcal{F} \colon \mathcal{H} \rightarrow \mathcal{H}$ defined by 
\bea \label{Def_F_sys}
[\mathcal{F} (\Psi) ](\theta) & := \begin{cases}
0, &  \theta \in[-\tau, 0),   \vspace{0.4em}\\ 
\mathbf{F} \Big(\Psi^S, \int_{-\tau}^0 \Psi^D(s) \d s\Big), & \theta = 0, 
\end{cases}  \quad \Forall \Psi = (\Psi^D, \Psi^S) \in  \mathcal{H},
\eea
is globally Lipschitz.

Then, for each $u_0 \in \mathcal{H}$, the mild solution of \eqref{Eq_DDE_Galerkin} emanating from $\Pi_N u_0$ converges uniformly to the mild solution of \eqref{Eq_abstract_ODE_DDE} emanating from $u_0$ on each bounded interval $[0, T]$, i.e.:
\be \label{uniform_conv_est}
\lim_{N\rightarrow \infty} \sup_{t \in [0, T]} \|u_N(t; \Pi_N u_0) - u(t; u_0)\|_{\mathcal{H}} = 0,  \qquad \Forall T > 0, \; u_0 \in \mathcal{H}.
\ee

\ec
\vspace{1ex}
\bp
This corollary is a direct consequence of Lemma~\ref{Lem_A1} and Lemma~\ref{Lem_A2}, ensuring respectively, Conditions {\bf (A1)} and {\bf (A2)} of Theorem~\ref{ParisVI_thm}. 
\ep
\vspace{1ex}

In the next two corollaries, we relax the global Lipschitz condition assumed in Corollary~\ref{Cor_DDE_global_Lip} to a local Lipschitz condition in addition to either a sublinear growth for $\mathcal{F}$ (see Corollary~\ref{Cor_DDE_local_Lip_Case1}) or an energy inequality satisfied by $\mathcal{F}$; see Corollary~\ref{Cor_DDE_local_Lip_Case2}.

%%%%%%%%%%%%%%%%%%%%%%BEGIN OF COROLLARY%%%%%%%%%%%%%%%%%%%
\bc \label{Cor_DDE_local_Lip_Case1}

Let  $\mathcal{A}$ be defined in \eqref{Def_A2} with $L_D$ specified in \eqref{Def_LD} and $L_S: \mathbb{R}^d \rightarrow \mathbb{R}^d$ to be a bounded linear operator. 
 
Assume that the nonlinearity $\mathcal{F}$ given by \eqref{Def_F_sys} is locally Lipschitz in the sense that for all $r > 0$ there exists $L(r)>0$ such that for any $\Psi_1$ and $\Psi_2$ in $\mathcal{H}$, we have
\be \label{Local_Lip_cond}
\|\Psi_1\|_{\mathcal{H}}<r \mbox{ and } \|\Psi_2\|_{\mathcal{H}} < r \Longrightarrow \|\mathcal{F}(\Psi_1) - \mathcal{F}(\Psi_2)\|_{\mathcal{H}} \le L(r) \|\Psi_1 - \Psi_2\|_{\mathcal{H}}.  
\ee

Assume also that $\mathcal{F}$ satisfies the following sublinear growth:
\be \label{Sublinear_onF}
\|\mathcal{F}(\Psi)\|_{\mathcal{H}} \le \gamma_1 \|\Psi\|_{\mathcal{H}} + \gamma_2, \qquad \Forall \Psi \in \mathcal{H},
\ee 
where $\gamma_1>0$ and $\gamma_2\geq 0$. 

Then, for each $u_0 \in \mathcal{H}$, the mild solution $u_N(t; \Pi_N u_0)$ of \eqref{Eq_DDE_Galerkin} emanating from $\Pi_N u_0$  and, the mild solution $u(t; u_0)$ of \eqref{Eq_abstract_ODE_DDE} emanating from $u_0$, do not blow up in any finite time. Moreover, $u_N(t; \Pi_N u_0)$ converges uniformly to $u(t; u_0)$ on each bounded interval $[0, T]$, i.e.:
\be \label{uniform_conv_est_Case1}
\lim_{N\rightarrow \infty} \sup_{t \in [0, T]} \|u_N(t; \Pi_N u_0) - u(t; u_0)\|_{\mathcal{H}} = 0,  \qquad \Forall T > 0, \; u_0 \in \mathcal{H}.
\ee
\ec

\bp

Recall that the local Lipschitz condition \eqref{Local_Lip_cond} on $\mathcal{F}$ ensures the existence and uniqueness of  a local mild solution $u(t; u_0)$ to \eqref{Eq_abstract_ODE_DDE} emanating from any $u_0 \in \mathcal{H}$; see e.g. \cite[Prop.~4.3.3]{Cazenave_al98}.\footnote{\cite[Prop.~4.3.3]{Cazenave_al98} is derived for the case of  contraction semigroups. However, the proof can be easily adapted to the case of more general $C_0$-semigroups $T(t)$ for which $\| T(t) \| \le M e^{\omega t}$.}

By recalling that  $\|T(t)\|_{\mathcal{H}} \le e^{\omega t}$ (see Lemma~\ref{Lem_A1}) and by using the sublinear growth assumption \eqref{Sublinear_onF} on $\mathcal{F}$, we obtain for mild solutions 
\bea \label{Energy_est_for_u}
\|u(t)\|_{\mathcal{H}} & \le \| T(t) u_0\|_{\mathcal{H}} + \int_{0}^t \|T(t-s)\mathcal{F}(u(s))\|_{\mathcal{H}} \d s  \\
& \le e^{\omega t} \|u_0\|_{\mathcal{H}} + \int_{0}^t e^{\omega(t-s)} \Big( \gamma_1 \|u(s)\|_{\mathcal{H}} + \gamma_2 \Big) \d s \\
& \le e^{\omega t} \|u_0\|_{\mathcal{H}} + \frac{\gamma_2  (e^{\omega t} -1)}{\omega} + \gamma_1  \int_{0}^t  e^{\omega(t-s)} \|u(s)\|_{\mathcal{H}}  \d s,
\eea 
where the positive constant $\omega$ is given by \eqref{omega}.

A simple multiplication by $e^{-\omega t}$ to both sides of \eqref{Energy_est_for_u}, leads then trivially to 
\bea
e^{-\omega t}\|u(t)\|_{\mathcal{H}} & \le \|u_0\|_{\mathcal{H}} + \frac{\gamma_2  (1 - e^{-\omega t})}{\omega} + \gamma_1  \int_{0}^t  e^{-\omega s} \|u(s)\|_{\mathcal{H}}  \d s \\
& \le \|u_0\|_{\mathcal{H}} + \frac{\gamma_2}{\omega} + \gamma_1  \int_{0}^t  e^{-\omega s} \|u(s)\|_{\mathcal{H}}  \d s.
\eea

An application of the Gronwall's inequality to $v(t):=e^{-\omega t}\|u(t)\|_{\mathcal{H}}$, gives then 
\be \label{Eq_bound_for_u}
\|u(t)\|_{\mathcal{H}} \le \Big( \|u_0\|_{\mathcal{H}} + \frac{\gamma_2}{\omega} \Big) e^{(\omega + \gamma_1) t},
\ee
preventing thus the blow up of a mild solution in finite time.

Similarly, for mild solutions $u_N$ of \eqref{Eq_DDE_Galerkin}, we have
\be \label{Eq_bound_for_uN}
\|u_N(t)\|_{\mathcal{H}} \le \Big( \|u_0\|_{\mathcal{H}} + \frac{\gamma_2}{\omega} \Big) e^{(\omega + \gamma_1) t}.
\ee
by noting that $\|\Pi_N\| < 1$ for all $N \ge 1$, $\|T_N(t)\|_{\mathcal{H}} \le e^{\omega t}$ (see Lemma~\ref{Lem_A1}), and by using the sublinear growth assumption \eqref{Sublinear_onF} on $\mathcal{F}$, preventing also  the blow up of any mild solution $u_N$ of \eqref{Eq_DDE_Galerkin},  in finite time.

Finally, \eqref{Eq_bound_for_u} and \eqref{Eq_bound_for_uN} lead to
\bea
& \|u(t; u_0)\|_{\mathcal{H}} \le C(T, \|u_0\|_{\mathcal{H}}), && \Forall t \in [0, T], \\
& \|u_N(t; \Pi_N u_0)\|_{\mathcal{H}} \le C(T, \|u_0\|_{\mathcal{H}}), && \Forall t \in [0, T] \text{ and } N \in \mathbb{N^*},
\eea
where
\bes
C(T, \|u_0\|_{\mathcal{H}}) := \Big( \|u_0\|_{\mathcal{H}} + \frac{\gamma_2}{\omega} \Big) e^{(\omega + \gamma_1) T}.
\ees 

Now, we can follow the proof of Theorem \ref{ParisVI_thm} to obtain the desired convergence result \eqref{uniform_conv_est_Case1}, with the only difference consisting of the global Lipschitz estimates used therein, by the local Lipschitz condition \eqref{Local_Lip_cond} applied to $\Psi$ such that 
\bes
\| \Psi\|_{\mathcal{H}}  \leq2 C(T, \|u_0\|_{\mathcal{H}}).
\ees

\ep
%%%%%%%%%%%%%%%%%%%%%%

%%%%%%%%%%%%%%%%%%%%%%BEGIN OF COROLLARY%%%%%%%%%%%%%%%%%%%
\bc \label{Cor_DDE_local_Lip_Case2}

Let  $\mathcal{A}$ be defined in \eqref{Def_A2} with $L_D$ specified in \eqref{Def_LD} and $L_S$ to be a bounded linear operator from $\mathbb{R}^d$ to $\mathbb{R}^d$. 
 
Assume that the nonlinearity $\mathcal{F}$ given by \eqref{Def_F_sys} is locally Lipschitz in the sense of \eqref{Local_Lip_cond}. Assume also that the  following energy inequality holds for $\mathcal{F}$
\be \label{Energy_ineq_onF}
\langle \mathcal{F}(\Psi), \Psi \rangle_{\mathcal{H}} \le \gamma_1 \|\Psi\|^2_{\mathcal{H}} + \gamma_2, \qquad \Forall \Psi \in \mathcal{H},
\ee 
where $\gamma_1 \in \mathbb{R}$ and $\gamma_2 \geq 0$. 

Then, for each $u_0 \in D(\mathcal{A})$, the strong solution $u_N(t; \Pi_N u_0)$ of \eqref{Eq_DDE_Galerkin} emanating from $\Pi_N u_0$,  and the strong solution\footnote{By strong solutions of \eqref{Eq_abstract_ODE_DDE}, we mean a solution in $C([0,T],D(\mathcal{A}))\cap C^1([0,T],\mathcal{H})$ of \eqref{Eq_abstract_ODE_DDE}.}  $u(t; u_0)$ of \eqref{Eq_abstract_ODE_DDE} emanating from $u_0$, do not blow up in any finite time. Moreover, $u_N(t; \Pi_N u_0)$ converges uniformly to $u(t; u_0)$ on each bounded interval $[0, T]$, i.e.:
\be \label{uniform_conv_est_engest_onF}
\lim_{N\rightarrow \infty} \sup_{t \in [0, T]} \|u_N(t; \Pi_N u_0) - u(t; u_0)\|_{\mathcal{H}} = 0,  \qquad \Forall T > 0, \; u_0 \in D(\mathcal{A}).
\ee

Furthermore, any strong solutions $v=u$ of \eqref{Eq_abstract_ODE_DDE} or $v=u_N$ of \eqref{Eq_DDE_Galerkin}, emanating respectively from $v(0)=u_0 \in D(\mathcal{A})$ or $v(0)=\Pi_N u_0$, have their $\mathcal{H}$-norm controlled as follows:
\be
 \|v(t)\|^2_{\mathcal{H}}\leq e^{\kappa t} \|v(0)\|^2_{\mathcal{H}} + \frac{2 \gamma_2}{\kappa} (e^{\kappa t}-1), \; t>0,
\ee
where $\kappa=2 (\omega+ \gamma_1)$, with $\omega$ given in  \eqref{omega}. 
\ec
%%%%%%%%%%%%%%%%%%%%%%END OF COROLLARY%%%%%%%%%%%%%%%%%%%

%%%%%%%%%%%%%%%%%%%%%%%%%%%%

%%%%%%%%%%%%%%%%%%%%%%%%%%%%%%%%%
\bp
Let $u_0\in D(\mathcal{A})$, and let $u$ be the mild solution of \eqref{Eq_abstract_ODE_DDE} emanating from $u_0$, such as ensured by the local Lipschitz condition on $\mathcal{F}$. Then by adapting the proof of e.g.~\cite[Prop.~4.3.9]{Cazenave_al98},\footnote{The regularity result  \cite[Prop.~4.3.9]{Cazenave_al98} is stated for the case of contraction semigroups. However, the proof can be adapted to the case of $C_0$-semigroups for which $\|T(t)\| \le e^{\omega t}$ (i.e. with $M=1$) such as encountered here when $L_D$ is as specified in \eqref{Def_LD}.}  we have that there exists a map $T:D(\mathcal{A})\rightarrow (0,\infty]$, for which $u\in C([0,T(u_0)),D(\mathcal{A}))\cap C^1([0,T(u_0)),\mathcal{H})$ and $u$ solves the initial-value problem
\begin{subequations}
\begin{eqnarray} \label{DDE_IVP}
&  \displaystyle{\frac{\d u}{\d t}} \hspace{-1.5em} & =\mathcal{A} u +\mathcal{F}(u), \label{DDE_IVP_eq1}\\
 & u(0)  \hspace{-0.5em}  & =u_0.
\end{eqnarray}
\end{subequations}

By taking the $\mathcal{H}$-inner product on both sides of \eqref{DDE_IVP_eq1} with the solution $u\in \mathcal{D}(\mathcal{A})$, and using the energy inequality \eqref{Energy_ineq_onF} and the stability property $\langle \mathcal{A} \Psi, \Psi \rangle_{\mathcal{H}} \le   \omega \| \Psi\|^2_{\mathcal{H}}$ from Lemma~\ref{Lemma_stability_prep}, we obtain
\be \label{Energy_est_for_u_2}
\frac{1}{2} \frac{\d}{\d t}\|u\|^2_{\mathcal{H}} = \langle \mathcal{A} u, u \rangle_{\mathcal{H}} + \langle \mathcal{F}(u), u \rangle_{\mathcal{H}} \le \omega \|u\|^2_{\mathcal{H}} +  \gamma_1 \|u\|^2_{\mathcal{H}} + \gamma_2,
\ee 
where the positive constant $\omega$ is given by \eqref{omega}.

It follows then from Gronwall's inequality that
\be \label{Eq_bounds_u_Case2}
\|u(t; u_0)\|^2_{\mathcal{H}} \leq e^{\kappa t} \|u_0\|^2_{\mathcal{H}} + \frac{2 \gamma_2}{\kappa} (e^{\kappa t}-1), \quad t \in [0, T(u_0)), \; u_0 \in D(\mathcal{A}),
\ee
where $\kappa = \omega + \gamma_1$.

Similarly, by noting that 
\bes
\langle \Pi_N \mathcal{F}(\Psi), \Psi \rangle_{\mathcal{H}}  = \langle \mathcal{F}(\Psi), \Psi \rangle_{\mathcal{H}}, \qquad \Forall \Psi \in \mathcal{H}_N, 
\ees
we have
\be
\|u_N(t; \Pi_N u_0)\|^2_{\mathcal{H}} \leq e^{\kappa t} \|u_0\|^2_{\mathcal{H}} + \frac{2 \gamma_2}{\kappa} (e^{\kappa t}-1), \quad t \in [0, T(u_0)), \; u_0 \in \mathcal{H}.
\ee

We have thus shown that for any initial data $u_0 \in D(\mathcal{A})$, the strong solutions $u(t; u_0)$  and $u_N(t; \Pi_N u_0)$ do not blow up in any finite time. The convergence result \eqref{uniform_conv_est_engest_onF}
can then be deduced as in the proof of Corollary~\ref{Cor_DDE_local_Lip_Case1}.

\ep
%%%%%%%%%%%%%%%%%%%%%%%%%%%%%%%%%

\vspace{1ex}
%%%%%%%%%%%%%%%%%%%%%%END OF COROLLARY%%%%%%%%%%%%%%%%%%%

\br \label{Rmk_forcing}
It is worth mentioning that the conclusions of Corollaries \ref{Cor_DDE_global_Lip}, \ref{Cor_DDE_local_Lip_Case1}  and \ref{Cor_DDE_local_Lip_Case2} still hold  when the underlying system of DDEs \eqref{Eq_nln_sys} is perturbed by a suitable time-dependent forcing, $\mathbf{g}(t)$. For instance, it suffices to assume that $\mathbf{g}(t) \in L^2_{\mathrm{loc}}([0,\infty); \mathbb{R}^d)$ to still get the convergence results. \qed
\er

\subsection{Examples}\label{Sec_examples}
In this section we provide some class of nonlinear scalar DDEs of the form \eqref{Eq_DDE} that fit with the assumptions of  Corollary \ref{Cor_DDE_local_Lip_Case2}.
In that respect we restrict our attention to the case of $\mathcal{A}$ such as defined in \eqref{Def_A} for $d=1$.  We discuss  below some classes of nonlinearities that verify the local Lipschitz condition \eqref{Local_Lip_cond} and the energy inequality  \eqref{Energy_ineq_onF}.
Extension to systems can be easily built out of these examples and are thus left to the reader. 

\subsubsection{Delay equations with a global Lipschitz nonlinearity}\label{sec_ex1}

 Let $\mathcal{F}$ be given such as in \eqref{Def_F}, and for which $F$ is assumed to be of the form
 \be\label{F_example0}
 F \Big(\Psi^S,\int_{-\tau}^0 \Psi^D(\theta) \d \theta \Big)=g\Big(\int_{-\tau}^0 \Psi^D(\theta) \d \theta\Big) +h(\Psi^S),
 \ee
 with $g$ and $h$ to be global Lipschitz maps from $\mathbb{R}$ to $\mathbb{R}$, of constants $L_1$ and $L_2$, respectively.  Then, we have 
  \bes
 \langle \mathcal{F}(\Psi), \Psi \rangle_{\mathcal{H}} =g \Big(\int_{-\tau}^0 \Psi^D(\theta) \d \theta \Big) \Psi^S +h(\Psi^S) \Psi^S, 
\ees
which gives
\beas
 \langle \mathcal{F}(\Psi), \Psi \rangle_{\mathcal{H}} &\leq L_1 \Big|\int_{-\tau}^0 \Psi^D(\theta) \d \theta\Big|\Big|\Psi^S\Big|+ (|g(0)|+|h(0)|)|\Psi^S| + L_2 |\Psi^S|^2\\
                        & \leq \gamma_1 \| \Psi \|^2_{\mathcal{H}} +\gamma_2,
\eeas
with $\gamma_1>0$ and $\gamma_2\geq 0$, and thus $F$ satisfies the energy inequality  \eqref{Energy_ineq_onF}. The local Lipschitz condition \eqref{Local_Lip_cond} for $\mathcal{F}$ is trivially satisfied  under the assumptions on $F$.  

Note that such nonlinear equations arise in many applications where a delayed monotone feedback mechanism is naturally involved in the description of the system's evolution; see \cite{GZT08,krisztin2008,Mallet_Sell96}.  It is also interesting to mention that such seemingly simple scalar DDEs with global Lipschitz nonlinearity can also support chaotic dynamics as illustrated in Section \ref{Sec_nearly-brownian} below.

\subsubsection{Delay equations with locally Lipschitz nonlinearity}\label{sec_ex2}
We relax now the global  Lipschitz requirement.   In that respect, we consider
\be \label{F_example}
F\Big(\Psi^S, \int_{-\tau}^0 \Psi^D(\theta) \d \theta \Big)=-  g_1 \Big(\int_{-\tau}^0 \Psi^D(\theta) \d \theta\Big) g_2\big(\Psi^S \big), 
\ee
and assume that 
\bi
\item[(i)] $g_1: \mathbb{R} \rightarrow \mathbb{R}^+$ is locally Lipschitz;

\item[(ii)] $g_2: \mathbb{R} \rightarrow \mathbb{R}$ is  locally Lipschitz and verifies the condition
\be\label{Eq_pos_g2}
g_2(x)x \ge 0, \qquad x \in \mathbb{R}.
\ee

\ei
These assumptions allow us to consider a  broad class of nonlinear effects that are not necessarily bounded or polynomial.

We check below that the abstract nonlinear map $\mathcal{F}$ defined  in  \eqref{Def_F} with $F$ given by \eqref{F_example} and that satisfy (i) and (ii),  satisfies also the conditions of  Corollary \ref{Cor_DDE_local_Lip_Case2}.  We first check the local Lipschitz condition.

Trivially, let us first remark that
\be \label{F_Lip_est0}
\|\mathcal{F}(\Psi_1) - \mathcal{F}(\Psi_2) \|_{\mathcal{H}} = \left |F \Big(\Psi_1^S, \int_{-\tau}^0 \Psi^D_1(\theta) \d \theta \Big) - F \Big(\Psi^S_2, \int_{-\tau}^0 \Psi^D_2(\theta) \d \theta \Big) \right|.
\ee

Let us introduce the notations $\alpha_i := \Psi_i^S$ and $\beta_i := \int_{-\tau}^0 \Psi^D_i(\theta) \d \theta$, $i = 1, 2$. Let $\Psi_i$ be chosen such that for $i = 1, 2$, $\|\Psi_i\|_{\mathcal{H}} \le R$ for some $R>0$. It follows that 
\be
|\alpha_i| \le R, \quad |\beta_i| = \left| \int_{-\tau}^0 \Psi^D_i(\theta) \d \theta \right| \le \sqrt{\tau} \|\Psi^D_i\|_{L^2},
\ee
and thus by definition of the $\mathcal{H}$-inner product \eqref{H_inner}
\be
 |\beta_i| \le \tau \|\Psi_i\|_{\mathcal{H}}  \le \tau R,
\ee
which leads to
\bea \label{F_Lip_est1}
& |F (\alpha_1, \beta_1) - F (\alpha_2, \beta_2) |\\
 & \le  |g_1(\beta_1) g_2(\alpha_1) - g_1(\beta_2) g_2(\alpha_2)| \\
 & \le  \Big|g_1(\beta_1) \big (g_2(\alpha_1) - g_2(\alpha_2) \big) \Big| +  \Big|\big( g_1(\beta_1) - g_1(\beta_2) \big) g_2(\alpha_2) \Big| \\
& \le  \mathrm{L}_2(R) |g_1(\beta_1)| |\alpha_1 - \alpha_2| + \mathrm{L}_1(\tau R) |g_2(\alpha_2)| |\beta_1 - \beta_2|,
\eea
where $\mathrm{L}_1(r)$ (resp.~$\mathrm{L}_2(r)$) denotes the local Lipschitz constant associated  with $g_1(x)$ (resp.~$g_2(x)$) for $|x|<r$.

On the other hand,
\beas
|g_1(\beta_1)| & \le |g_1(\beta_1) - g_1(0)| + |g_1(0)|  \\
& \le \mathrm{L}_1(\tau R) |\beta_1| +  |g_1(0)| \\
& \le \tau  R \mathrm{L}_1(\tau  R)  +  |g_1(0)|,
\eeas
and
\beas
|g_2(\alpha_2)| \le  R \mathrm{L}_2(R) +  |g_2(0)|.
\eeas
Note also that 
\bes
|\alpha_1 - \alpha_2| \le \| \Psi_1 - \Psi_2 \|_{\mathcal{H}}, 
\ees
and that
\bes
|\beta_1 - \beta_2| \le \sqrt{\tau} \|\Psi^D_1 - \Psi^D_2\|_{L^2} \le \tau  \|\Psi_1 - \Psi_2\|_{\mathcal{H}}. 
\ees
We obtain then from \eqref{F_Lip_est1} that 
\be \label{F_Lip_est2}
|F (\alpha_1, \beta_1) - F (\alpha_2, \beta_2) | \le L(R)  \|\Psi_1 - \Psi_2\|_{\mathcal{H}},
\ee
where
\bes 
L(R):= \mathrm{L}_2(R) \Big(\sqrt{\tau}  R \mathrm{L}_1(\tau R)  +  |g_1(0)| \Big) 
+ \tau \mathrm{L}_1(\tau R)  \Big(R \mathrm{L}_2( R) +  |g_2(0)| \Big),
\ees
which gives the local Lipschitz property of $\mathcal{F}$ as a map from $\mathcal{H}$ to $\mathcal{H}$.

The energy inequality \eqref{Energy_ineq_onF} is here readily satisfied since
\be
\langle \mathcal{F}(\Psi), \Psi \rangle_{\mathcal{H}} = F \Big(\Psi^S, \int_{-\tau}^0 \Psi^D(\theta) \d \theta \Big) \Psi^S \leq 0,
\ee
because of \eqref{Eq_pos_g2}.

%%%%%%%%%%%%%%%%%%%%
\needspace{1\baselineskip}
\br\label{Rmk_ex}
\hspace*{2em}  \vspace*{-0.4em}
\bi
\item[(i)] Note that famous delayed models from population dynamics are covered by Corollary \ref{Cor_DDE_local_Lip_Case2}, although not satisfying \eqref{F_example}. For instance, delayed logistic equations of the form 
\be
\frac{\d x}{\d t}=r x(t)\Big(1-K^{-1}\int_{t-\tau}^t \omega(s) f(x(s))\d s\Big), \; r, K, \tau>0,
\ee
with $\omega \in L^{\infty} (\mathbb{R},\mathbb{R}^+)$ and $f\in L^1(\mathbb{R},\mathbb{R}^+)$ satisfying the inequality 
\bes
 |f(x)-f(y)| \leq \gamma |x-y|, 
\ees
for some $\gamma > 0$ and for almost every $x,y \in \mathbb{R}$, are still covered by Corollary \ref{Cor_DDE_local_Lip_Case2}. 

\item[(ii)] Note also that many other nonlinear effects could have been considered in  \eqref{Eq_DDE} for which the convergence result of Corollary \ref{Cor_DDE_local_Lip_Case2} would hold. For instance we could have  considered 
\be\label{other_nonl}
F\Big(\Psi^S, \int_{-\tau}^0 \Psi^D(\theta) \d \theta\Big)= \sum_{j=1}^{2p-1} b_j \Big(\int_{-\tau}^0 \Psi^D(\theta) \d \theta\Big) (\Psi^S)^j, p\geq 2,
\ee
where each $b_j$ is a local Lipschitz  function in $L^{\infty}(\mathbb{R})$ and $b_{2p-1} (\cdot )\leq \beta <0$ for some constant $\beta$. 

Under these assumptions, the nonlinearity \eqref{other_nonl} satisfies the energy inequality \eqref{Energy_ineq_onF} with $\gamma_1 = 0$ and with $\gamma_2$ sufficiently large, depending on the $L^\infty$ norm of
the functions $b_j$'s. This is because the term with the largest power $(\Psi^S)^{2p-1}$ is strictly 
negative by  assumption on $b_{2p-1}$, and the other terms with a lower degree can be controlled by using Young's inequality.  \qed

\ei
\er

%%%%%%%%%%%%%%%%%%%%%%%%%%
\section{Galerkin approximation: Analytic formulas for scalar DDEs}\label{Sec_Galerkin_analytic}

This section is devoted to the derivation of explicit expressions of the Galerkin approximation \eqref{Eq_DDE_Galerkin} associated with 
nonlinear DDEs. For simplicity, we focus on the case of a scalar DDE taking the form given by \eqref{Eq_DDE}. The more general case of nonlinear systems of DDEs is dealt with in Appendix \ref{Appendix_systems}; see also Appendix \ref{Subsect_var_C} for the case where the linear part of   \eqref{Eq_DDE} involves  a  distributed-delay term as in \eqref{Eq_nln_sys}. 

As a preparation for the forthcoming analytic derivations, we need to express the derivative of the Koornwinder polynomials in terms of the polynomials themselves. This is the content of the following proposition.

\begin{prop} \label{prop:dPn}
The Koornwinder polynomial $K_n$ of degree $n\in \mathbb{N}$ defined in \eqref{eq:Pn} satisfies the differential relation
\be \label{eq:dPn}
\frac{\d K_n}{\d s}(s) = \sum_{k = 0}^{n-1} a_{n,k} K_k(s), \quad s \in (-1,1),
\ee
where the coefficients $\boldsymbol{a}_n:=(a_{n,0}, \cdots, a_{n,n-1})^\tr$, satisfy the upper triangular system of linear equations
\be \label{eq:algebraic}
\mathbf{T}\boldsymbol{a}_n = \boldsymbol{b}_n,
\ee
with $\mathbf{T}:=(\mathbf{T}_{i,j})_{n\times n}$ and $b_{n}:=(b_{n,0}, \cdots, b_{n,n-1})^\tr$ given by
\bea \label{eq:algebraic_def}
\mathbf{T}_{i,j} & = \begin{cases}
0, & \; \text{ if } j < i,\\
i^2 + 1, & \; \text{ if } j = i,\\
-(2i+1), & \; \text{ if } j > i,
\end{cases} \; \qquad \text{ where } \qquad 0 \le i, j \le n-1, \\
b_{n,i} & = \begin{cases}
\frac{1}{2}(2i+1)(n+i+1)(n-i), & \text{ if $n+i$ is even}, \vspace{1em}\\
(n^2 + n)(2i+1) - \frac{i}{2}(n+i)(n-i+1) & \\
\hspace{2em} -\frac{ i}{2}(i+1)(n-i-1)(n+i+2), & \text{ if $n+i$ is odd}.
\end{cases}
\eea

For the rescaled version $K^\tau_n$ defined by \eqref{eq:Pn_tilde}, it holds that
\be \label{eq:dKn_tau}
\frac{\d K^\tau_n}{\d \theta}(\theta) = \frac{2}{\tau} \sum_{k = 0}^{n-1} a_{n,k} K^\tau_k(\theta), \qquad n \in \mathbb{N}, \quad \theta\in (-\tau,0).
\ee

\end{prop}

\bp

See Appendix \ref{sect:coef_matrix_proof}.

\ep

Let us now rewrite the unknown $u_N$ in the Galerkin system \eqref{Eq_DDE_Galerkin} in terms of the first $N$ rescaled Koornwinder polynomials, i.e.:
\be \label{x_t expand}
u_N(t) = \sum_{n=0}^{N-1} y_n(t) \mathcal{K}^\tau_n, \qquad t \ge 0,
\ee
where
\be
y_n(t) = \frac{\langle u_N(t), \mathcal{K}^\tau_n \rangle_{\mathcal{H}}}{\| \mathcal{K}^\tau_n \|^2_{\mathcal{H}}}.
\ee

We then replace $u_N$ in Eq.~\eqref{Eq_DDE_Galerkin} by the expansion given in \eqref{x_t expand}, and take the $\mathcal{H}$-inner product on both sides with $\mathcal{K}^\tau_j$ for each $j \in \{0, \cdots, N-1
\}$ to obtain:
\be \label{GalerkinCalc_v1}
\|\mathcal{K}^\tau_j\|_{\mathcal{H}}^2 \frac{\d y_j}{\d t} = \sum_{n=0}^{N-1} y_n(t) \left \langle \mathcal{A}_N \mathcal{K}^\tau_n, \mathcal{K}^\tau_j \right \rangle_{\mathcal{H}} + \Bigg \langle \Pi_N \mathcal{F} \Bigg ( \sum_{n=0}^{N-1} y_n(t) \mathcal{K}^\tau_n \Bigg), \mathcal{K}^\tau_j \Bigg \rangle_{\mathcal{H}}.
\ee

Recall that the linear operator $\mathcal{A}$ here is defined by \eqref{Def_A}. Then, for each $n \in \{0, \cdots, N-1
\}$, it holds that
\bea
\mathcal{A}_N \mathcal{K}^\tau_n = \Pi_N \mathcal{A} \mathcal{K}^\tau_n & = \sum_{l = 0}^{N-1} \Big(\frac{1}{\tau} \Big \langle \frac{\d^+}{\d \theta} K_n^\tau, K_l^\tau \Big \rangle_{L^2} \\
& \qquad + \Big(a K^\tau_n(0) + b K_n^\tau(-\tau) + c \int_{-\tau}^0 K_n^\tau(\theta) \d
\theta \Big) \Big) \frac{\mathcal{K}_l^\tau }{\|\mathcal{K}_l^\tau \|_{\mathcal{H}}^2}.
\eea

We obtain then that
\bea \label{GalerkinCalc_part1_1}
\left \langle \mathcal{A}_N \mathcal{K}^\tau_n, \mathcal{K}^\tau_j \right \rangle_{\mathcal{H}} & = \frac{1}{\tau} \Big \langle \frac{\d^+}{\d \theta} K_n^\tau, K_j^\tau \Big \rangle_{L^2} \\
& \quad + \Big( a K^\tau_n(0) + b K^\tau_n(-\tau) + c \int_{-\tau}^0 K_n^\tau(\theta) \d \theta\Big) K^\tau_j(0).
\eea

It follows from the expression of $\frac{\d K^\tau_n}{\d \theta}$ in \eqref{eq:dKn_tau} that
\bea \label{GalerkinCalc_part1_2}
\frac{1}{\tau} \int_{-\tau}^0 \frac{\d^+ K^\tau_n}{\d \theta}(\theta) K^\tau_j(\theta)\d \theta & = \frac{2}{\tau}\sum_{k=0}^{n-1} a_{n,k} \left( \frac{1}{\tau} \int_{-\tau}^0 K^\tau_k(\theta) K^\tau_j(\theta)\d \theta \right) \\
& = \frac{2}{\tau}\sum_{k=0}^{n-1} a_{n,k} \left( \langle \mathcal{K}^\tau_k, \mathcal{K}^\tau_j \rangle_{\mathcal{H}} - K^\tau_k(0) K^\tau_j(0) \right ) \\
& = \frac{2}{\tau} \sum_{k=0}^{n-1} a_{n,k} \left( \delta_{j,k} \|\mathcal{K}^\tau_j\|^2_{\mathcal{H}} - 1 \right ),
\eea
where $\delta_{j,k}$ denotes the Kronecker delta, and the last equality above follows from the orthogonal property of the Koornwinder polynomials  as well as the normalization property $K^\tau_n(0) = 1$; cf.~\eqref{Eq_normalization}.

Note that $K^\tau_0 \equiv 1$, which follows from the definition of the Koornwinder polynomials $K_n$ given by \eqref{eq:Pn} and the fact that the first Legendre polynomial $L_0$ is identically $1$. We get then $\|\mathcal{K}^\tau_0\|^2_{\mathcal{H}} = 2$. Note also that
\be
\langle \mathcal{K}^\tau_n, \mathcal{K}^\tau_0 \rangle_{\mathcal{H}} =
\frac{1}{\tau} \int_{-\tau}^0 K^\tau_n(\theta) \d \theta + 1 = \delta_{n,0} \|\mathcal{K}^\tau_0\|^2_{\mathcal{H}} = 2 \delta_{n,0},
\ee
leading thus to
\be \label{Eq_int_Kn}
\int_{-\tau}^0 K^\tau_n(\theta) \d \theta = \tau (2 \delta_{n,0} - 1), \quad n \in \mathbb{N}.
\ee
By using \eqref{Eq_int_Kn}, the normalization property $K^\tau_j(0) = 1$ and the identity $K^\tau_j(-\tau) = K_j(-1)$ (valid for any $j \ge 0$), we obtain
\be \label{GalerkinCalc_part1_3}
\Big( a K^\tau_n(0) + b K^\tau_n(-\tau) + c \int_{-\tau}^0 K_n^\tau(\theta) \d \theta \Big) K^\tau_j(0) = a + b K_n(-1) + c \tau (2 \delta_{n,0} - 1).
\ee

Now, by using \eqref{GalerkinCalc_part1_2} and \eqref{GalerkinCalc_part1_3} in \eqref{GalerkinCalc_part1_1}, we obtain
\bea \label{GalerkinCalc_part1}
\sum_{n=0}^{N-1} y_n(t) \left \langle \mathcal{A}_N \mathcal{K}^\tau_n, \mathcal{K}^\tau_j \right \rangle_{\mathcal{H}} & = \sum_{n=0}^{N-1} \Bigl ( a + b K_n(-1) + c \tau (2 \delta_{n,0} - 1) \\
& \hspace{5em} + \frac{2}{\tau} \sum_{k=0}^{n-1} a_{n,k} \left( \delta_{j,k} \|\mathcal{K}_j\|^2_{\mathcal{H}} - 1 \right ) \Bigr) y_n(t).
\eea

For the nonlinear part, since $\langle \Pi_N \Phi, \mathcal{K}^\tau_n \rangle_{\mathcal{H}} = \langle \Phi, \mathcal{K}^\tau_n \rangle_{\mathcal{H}}$ for all $\Phi \in \mathcal{H}$ and all $n \in \{0, \cdots, N-1\}$, together with the definition of $\mathcal{F}$ given in \eqref{Def_F}, we obtain
\be \label{GalerkinCalc_part2a}
\Bigl \langle \Pi_N \mathcal{F} \Bigl ( \sum_{n=0}^{N-1} y_n(t) \mathcal{K}^\tau_n \Bigr ), \mathcal{K}^\tau_j \Bigr \rangle_{\mathcal{H}} = F \Biggl( \sum_{n=0}^{N-1} y_n(t), \int_{-\tau}^0 \sum_{n=0}^{N-1} y_n(t) K^\tau_n(\theta) \d \theta \Biggr).
\ee
From \eqref{Eq_int_Kn}, it also follows that
\be \label{GalerkinCalc_part2b}
\int_{-\tau}^0 \sum_{n=0}^{N-1} y_n(t) K^\tau_n(\theta) \d \theta =
\tau y_0(t) - \tau \sum_{n=1}^{N-1} y_n(t).
\ee
By using \eqref{GalerkinCalc_part2b} in \eqref{GalerkinCalc_part2a}, we get
\be \label{GalerkinCalc_part2}
\Bigl \langle \Pi_N \mathcal{F} \Bigl ( \sum_{n=0}^{N-1} y_n(t) \mathcal{K}^\tau_n \Bigr ), \mathcal{K}^\tau_j \Bigr \rangle_{\mathcal{H}} = F \Biggl( \sum_{n=0}^{N-1} y_n(t),
\tau y_0(t) - \tau \sum_{n=1}^{N-1} y_n(t) \Biggr).
\ee

Now, by using \eqref{GalerkinCalc_part1} and \eqref{GalerkinCalc_part2} and recalling that $\|\mathcal{K}^\tau_n\|_{\mathcal{H}} = \|\mathcal{K}_n\|_{\mathcal{E}}$, we obtain from Eq.~\eqref{GalerkinCalc_v1} the following explicit form of the $N$-dimensional Galerkin system \eqref{Eq_DDE_Galerkin}:
\begin{equation} \label{Galerkin_AnalForm}
\begin{aligned}
\frac{\d y_j}{\d t} & = \frac{1}{\|\mathcal{K}_j\|_{\mathcal{E}}^2 } \sum_{n=0}^{N-1} \Big( a + b K_n(-1) + c \tau (2 \delta_{n,0} - 1) \\
& \hspace{8em} + \frac{2}{\tau}\sum_{k=0}^{n-1} a_{n,k} \left( \delta_{j,k} \|\mathcal{K}_j\|^2_{\mathcal{E}} - 1 \right) \Big) y_n(t) \\
& \hspace{1em} + \frac{1}{\|\mathcal{K}_j\|_{\mathcal{E}}^2} F \left( \sum_{n=0}^{N-1} y_n(t), \tau y_0(t) - \tau \sum_{n=1}^{N-1} y_n(t) \right), \;\; 0\leq j\leq N-1.
\end{aligned}
\end{equation}

For later usage, we rewrite the above Galerkin system into the following compact form:
\be \label{Galerkin_cptForm}
\boxed{\frac{\d \boldsymbol{y}}{\d t} = A \boldsymbol{y} + G (\boldsymbol{y}),}
\ee
where $A \boldsymbol{y}$ denotes the linear part of Eq.~\eqref{Galerkin_AnalForm}, and $G(\boldsymbol{y})$ the nonlinear part. Namely, $A$ is the $N\times N$ matrix whose elements are given by
\begin{equation} \label{eq:A}
\boxed{
\begin{aligned}
(A)_{j,n} & = \frac{1}{\|\mathcal{K}_j\|_{\mathcal{E}}^2 }\Big(a + b K_n(-1) + c \tau (2 \delta_{n,0} - 1) \\
& \hspace{8em}+ \frac{2}{\tau}\sum_{k=0}^{n-1} a_{n,k} \left( \delta_{j,k} \|\mathcal{K}_j\|^2_{\mathcal{E}} - 1 \right ) \Big),
\end{aligned}
}
\end{equation}
where $j, n = 0, \cdots, N-1$, and the nonlinear vector field $G \colon \mathbb{R}^N \rightarrow \mathbb{R}^N$, is given component-wisely
by
\be \label{eq:G}
\boxed{G_j(\boldsymbol{y}) = \frac{1}{\|\mathcal{K}_{j}\|_{\mathcal{E}}^2} F \left( \sum_{n=0}^{N-1} y_n(t), \tau y_0(t) - \tau \sum_{n=1}^{N-1} y_n(t) \right), \; 0\leq j\leq N-1.}
\ee

\medskip
\br  \label{Rmk_Galerkin_forcing}

When a time-dependent forcing $g(t)$ is added to the RHS of the DDE \eqref{Eq_DDE}, the only change in the corresponding Galerkin system is to add a term $\|\mathcal{K}_{j}\|_{\mathcal{E}}^{-2} g(t)$ to each $G_j$ in \eqref{eq:G}. Recall also from Remark~\ref{Rmk_forcing} that it is sufficient to require that $g\in L_{\mathrm{loc}}^2(\mathbb{R}; \mathbb{R})$ in order to ensure the convergence result of the Galerkin system over any finite interval. \qed

\er

%%%%%%%%%%%%%%%%%%%%%%%%%%%%%%%%%%%%%%%%%%%%%%%%
\section{Approximation of chaotic dynamics: Numerical results} \label{Sect_Numerics}
In this section, we report on the performance of our Galerkin schemein approximating quasi-periodic and 
chaotic dynamics. 
In the case of  the latter, it is well known that any finite-time uniform convergence result ---  such as the one given by \eqref{uniform_conv_est_engest_onF} and obtained in Corollary \ref{Cor_DDE_local_Lip_Case2} --- becomes less useful, due to sensitivity to initial data. 

 In this case, when individual solutions diverge exponentially, it is natural to consider instead the approximation of the strange attractor or of the statistics of the dynamics. More generally, the approximation of meaningful invariant probability measures supported by the strange attractor  is of primary interest in chaotic dynamics. Nonlinear DDEs, like those considered here, are known to support such probability measures once a global attractor is known to exist \cite{chekroun_glatt-holtz}.

If, in addition to the finite-time uniform convergence \eqref{uniform_conv_est_engest_onF}, 
a uniform dissipativity assumption is satisfied,  then any sequence of invariant measures associated with the Galerkin approximation converges weakly to an invariant measure of the full system; see \cite[Thm.~2.2]{wang2009approximating}. 
The uniform dissipativity assumption referred to in \cite{wang2009approximating} is satisfied if one can establish, uniformly in $N$, the existence of  an absorbing ball for the Galerkin reduced systems in another separable Hilbert space $\mathcal{V}$, which is compactly imbedded in $\mathcal{H}$, in the case of strongly dissipative systems.

We show in Sections~\ref{Sec_nearly-brownian} and \ref{Sect_bimodal}
below that, for two simple nonlinear scalar DDEs that --- even in instances in which the 
 uniform dissipativity assumption  of \cite[Thm.~2.2]{wang2009approximating} is not guaranteed --- our Galerkin scheme is still able to approximate significant statistical properties of the chaotic dynamics, or the strange attractor itself.  Finally, we illustrate in Section~\ref{Sect_ENSO}, {\mg for a highly idealized ENSO model from climate dynamics} that our approach also works in the case of periodically forced DDEs with multiple delays.

\subsection{``Nearly-Brownian''  chaotic dynamics.}\label{Sec_nearly-brownian}
We consider here a modified version of the  DDE analyzed in \cite{sprott2007simple} with a sinusoidal nonlinearity.
The modification consists of  replacing the discrete delay by distributed delays. As we will see, this modification does not affect the main dynamical properties identified in \cite{sprott2007simple} in the case of a discrete delay, once the proper parameter values are chosen. 

More precisely, we consider the following DDE
 \be\label{Eq_sindel}
\dot{x}=a\sin\Big(\int_{t-\tau}^tx(s) \d s\Big),
\ee
with $a=0.5$ and $\tau=5.5$. This example fits within the general class discussed in Section \ref{sec_ex1}, to which the rigorous convergence results described in Sections \ref{Subsect_DDE_Galerkin} in an abstract setting,  and in Section \ref{Sec_Galerkin_analytic} more concretely, do apply.

As pointed out in the introduction of this section, such finite-time uniform convergence results 
are essential in general, but they are not the ones we are necessarily looking for 
 in approximating chaotic dynamics. We rely, therefore, on careful numerical simulations
to explore the performance of our Koornwinder-polynomial--based Galerkin systems to approximate the statistical features of the chaotic dynamics in these examples. 

\begin{figure}[hbtp]
   \centering
\includegraphics[height=0.4\textwidth,width=.9\textwidth]{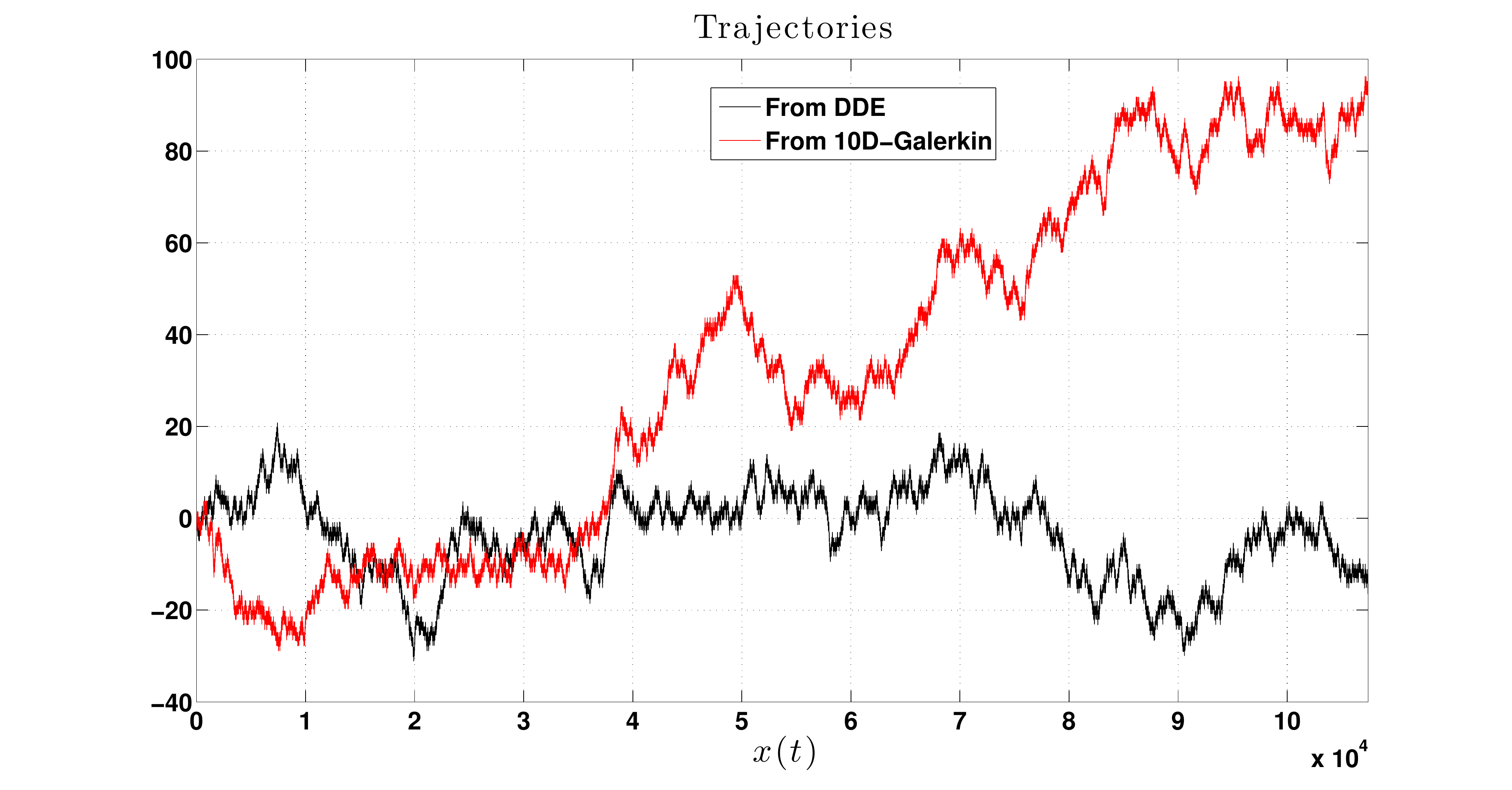}
  \caption{{\footnotesize {\mg Trajectory  $x(t)$ simulated by} the DDE \eqref{Eq_sindel} (black curve), and its 10D-Galerkin approximation {\mg $x_N(t)$, with $N = 10$} (red curve), obtained by the method described in Section~\ref{Sec_Galerkin_analytic}.}}   \label{Trajectories_sindel}
\end{figure}

A sample trajectory of Eq.~\eqref{Eq_sindel} is shown in black in Fig.~\ref{Trajectories_sindel}. While the governing DDE is perfectly deterministic, the visual resemblance of this trajectory to a sample path of Brownian motion is obvious. A trajectory with the same constant initial history over the interval $[-\tau, 0)$, but obtained by solving a 10D-approximation by our Galerkin scheme of the DDE, is shown as the red curve in the figure. It is clear that individual trajectories do have the same overall behavior, which is nearly Brownian, but the pointwise approximation of the exact trajectory by the approximate one is not good.

\begin{figure}[hbtp]
   \centering
\includegraphics[height=0.7\textwidth,width=.9\textwidth]{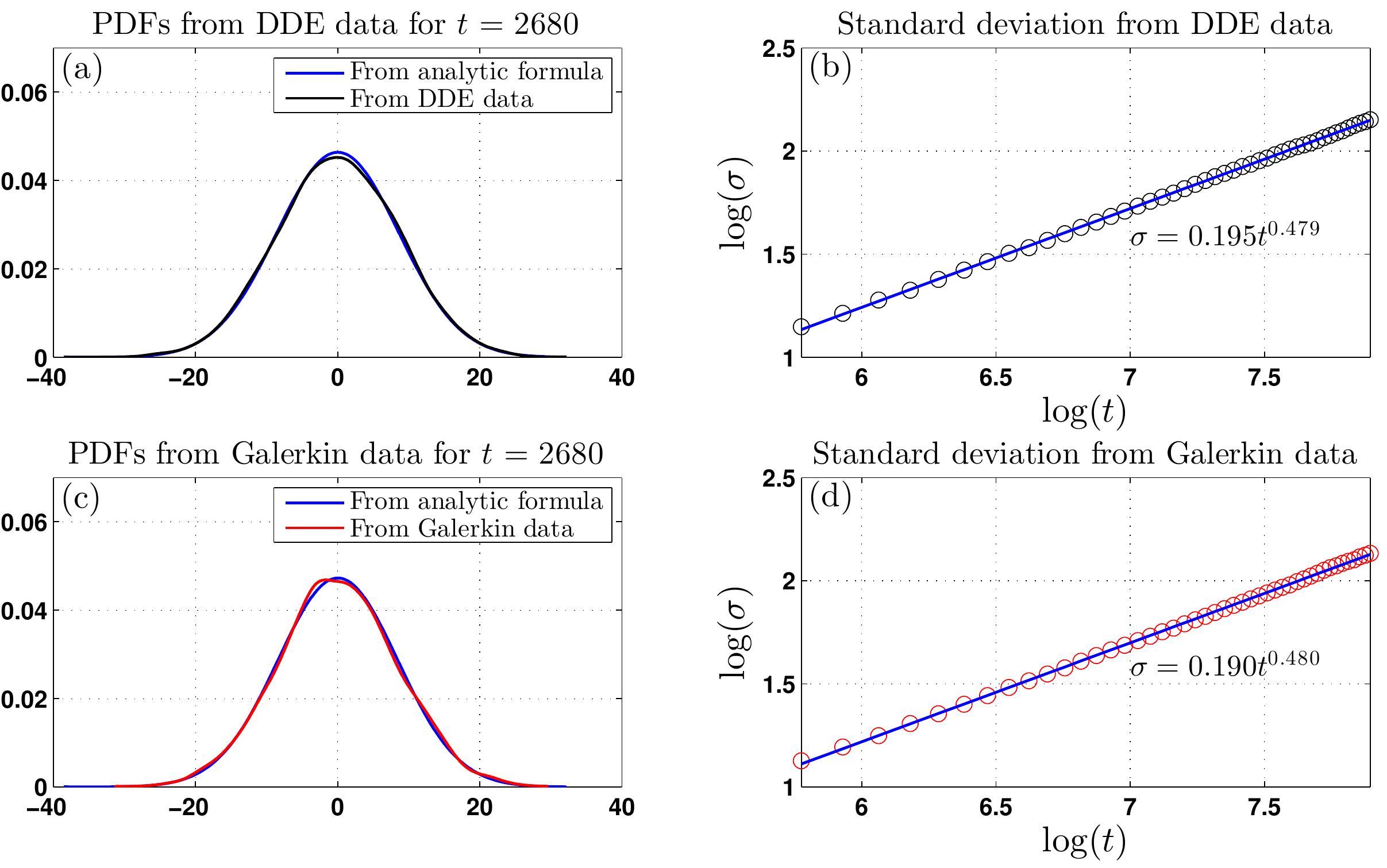}
  \caption{{\footnotesize {\mg (a, c):} Probability density functions (PDFs); and (b, d): standard deviations $\sigma$, as estimated from the simulations of the DDE \eqref{Eq_sindel} (black curve {\mg and black circles, in the left panels and the right panels, respectively),} and its 10D-Galerkin approximation (red curve {\mg and red circles)} by the method described in Section~\ref{Sec_Galerkin_analytic}, respectively. In each of the panels (a) and (b), the results are compared with a Gaussian distribution estimated by standard analytic formulas (blue curves), and in each of the panels (c) and (d), the results are compared with the best linear regression fit (blue curve) of $\log(\sigma)$ versus $\log(t)$,  providing the corresponding slope and the exponent reported therein.}}   \label{Sine_model_statistics}
\end{figure}

To study instead the statistics of the solutions,
we have estimated ---  for a collection of $n=10^4$ initial histories of constant value drawn uniformly in $[-1,1]$ --- the empirical probability
density  function (PDF) of the corresponding $x(t)$-values at a given time $t$. 
Figure \ref{Sine_model_statistics} reports on the numerical results at $t=2680$ when $x(t)$ is simulated from Eq.~\eqref{Eq_sindel} by forward Euler integration (black curve) with a time step of $\Delta  t = \tau/2^{10}$, and when $x(t)$ is approximated  by a 10D-Galerkin approximation (red curve)  obtained from the analytic formulas \eqref{Galerkin_cptForm}--\eqref{eq:G} applied to Eq.~\eqref{Eq_sindel}.  

The Galerkin system {\mg of ODEs} is integrated using a semi-implicit Euler method 
that still uses $\Delta  t = \tau/2^{10}$, but in which the linear part $A y$ is treated implicitly, {\mg while} the nonlinear term $G(y)$ is treated explicitly.  The approximate solution $x_N(t)$, provided by an $N$-dimensional Galerkin system of the form \eqref{Galerkin_cptForm}, is obtained by using the expansion \eqref{x_t expand} into Koornwinder polynomials. More precisely, 
$x_N(t)$ is obtained as the state part of $u_N$ given by  \eqref{x_t expand} which, thanks to the normalization property $K_n^\tau(0)  = 1$ given in ~\eqref{Eq_normalization}, reduces to
\be
x_N(t) = \sum_{j = 1}^N y_j(t),
\ee 
where $y:=(y_0, \cdots, y_{N-1})$  is the vector solution of \eqref{Galerkin_cptForm}. 

Also shown in blue in panels (a, c) of Fig.~\ref{Sine_model_statistics}, is a Gaussian distribution with the same mean and standard deviation, both of which are estimated from the $x(t)$-values at $t=2680$ of the simulated DDE solutions. In both cases, the empirical distributions, as obtained from  the simulated  DDE solution  or as obtained by using a 
Galerkin approximation with $N=10$, closely follow a Gaussian law. 

For both $x(t)$, as simulated by numerically solving Eq.~\eqref{Eq_sindel}, and $x_N(t)$, as obtained by using a 
Galerkin approximation with $N=10$, the standard deviations of the corresponding collection of trajectories
are shown versus time in panels (b) and (d) of Fig.~\ref{Sine_model_statistics}. The best linear regression fit of $\log(\sigma)$ versus $\log(t)$ gives $\sigma=0.195 t^{0.479}$, in the case of Eq.~\eqref{Eq_sindel}, and $\sigma=0.190 t^{0.480}$, in the case of the 10D-Galerkin approximation.

In both cases, the slopes of the fitted lines indicate that the {\mg deterministic dynamics of Eq.~\eqref{Eq_sindel}} mimics that of Brownian motion, for which the slope would be $0.5$, with a diffusion
coefficient $D=\sigma^2/t$. In particular, the solutions do not stay within any bounded subset; the dynamics thus violates any dissipation criterion, while still possessing an attractor with strange behavior (not shown). 

To the best of our knowledge, the 10D-Galerkin approximation computed here {\mg thus provides} 
the first example of a chaotic system of ODEs whose solutions exhibit a statistical behavior close to that of Brownian motion.  

%%%%%%%%%%%%%%%%%%%%%%%%%%

\subsection{Bimodal chaotic dynamics with low-frequency variability.} \label{Sect_bimodal}
The model studied in this subsection,
\be\label{Eq_DDE_chafee}
\dot{x}=a x(t-\tau) -b x(t-\tau)^3,
\ee
is also based on \cite{sprott2007simple}; the parameter values used here are $a=0.5$, $b=20$ and $\tau=3.35$.
 Remark~\ref{Rmk_Chafee} at the end of this subsection discusses similarities between the dynamics of the model above and important aspects of ENSO dynamics.

\begin{figure}[hbtp]
   \centering
\includegraphics[height=0.5\textwidth,width=.9\textwidth]{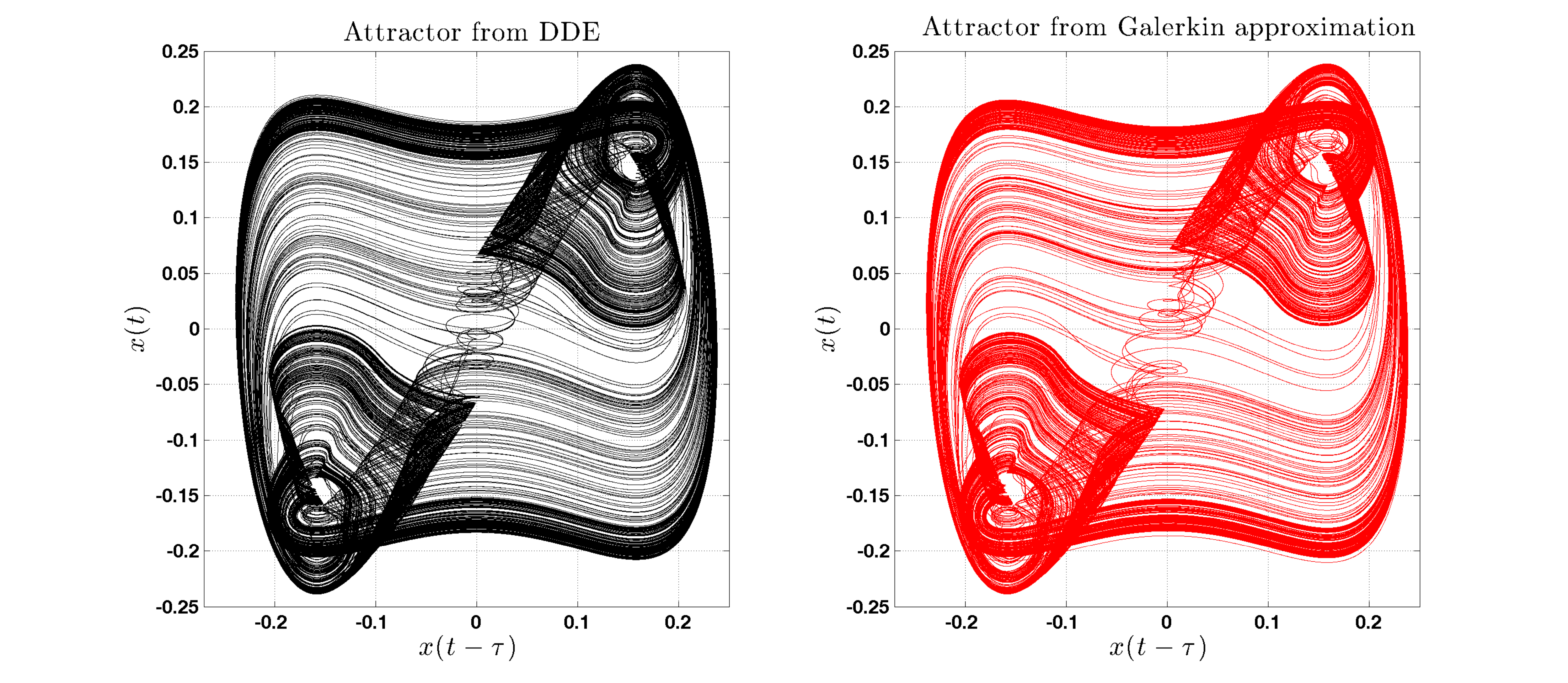}
  \caption{{\footnotesize The attractor associated with Eq.~\eqref{Eq_DDE_chafee} (left panel) and its approximation obtained from a 6D-Galerkin approximation (right panel).}} \label{Super_chafee} 
\end{figure}

\begin{figure}[hbtp]
   \centering
\includegraphics[height=0.4\textwidth,width=.8\textwidth]{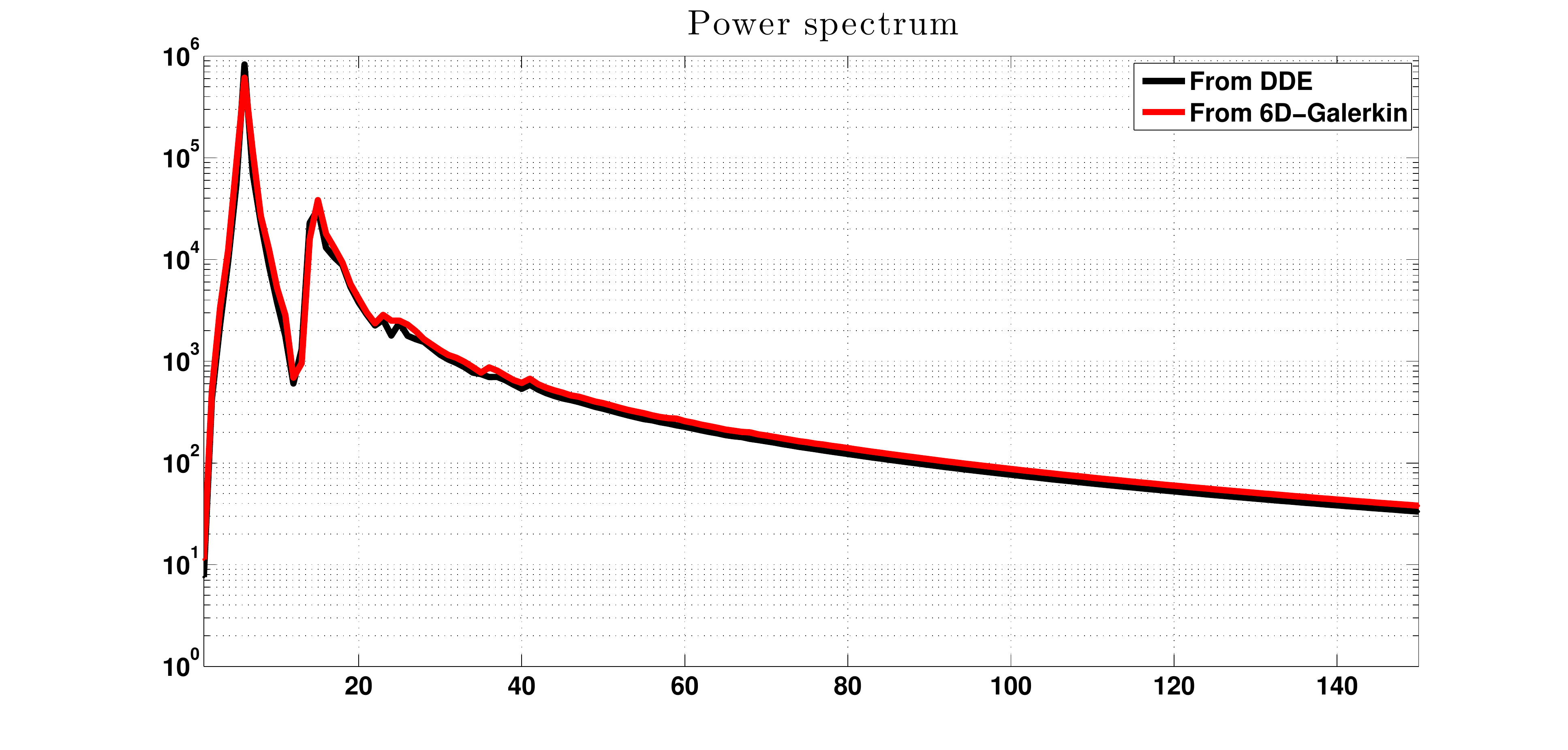}
  \caption{{\footnotesize Power spectral density of $x(t)$ as simulated from Eq.~\eqref{Eq_DDE_chafee}, and from a 6D-Galerkin approximation derived from the analytic formulas described in Section \ref{Sec_Galerkin_analytic} and applied to Eq.~\eqref{Eq_DDE_chafee}.}}   \label{PSD_chafee}
\end{figure}

In contrast to the DDE considered in the previous {\mg subsection,} Eq.~\eqref{Eq_DDE_chafee} does not fit directly within the general framework of Section \ref{Subsect_DDE_Galerkin}, for which rigorous convergence results are available.
The discrete lag present in the cubic term leads to complications for a rigorous analysis. Replacing this lag effect by a distributed one, as in Eq.~\eqref{Eq_sindel},  would place the DDE of Eq.~\eqref{Eq_DDE_chafee} into the class considered in Section~\ref{sec_ex2}, for which finite-time uniform convergence results are guaranteed. 
But even then, we cannot be assured {\it a priori} of an effective approximation of statistical features of the dynamics, as discussed above.  

The purpose of this subsection is to show that, even in such a borderline case with  respect to
the theory presented in this article,  statistical and even topological features can still be
remarkably well approximated by the Galerkin systems of Section \ref{Sec_Galerkin_analytic}, when appropriately modified to handle the case of discrete delay in the nonlinear terms.\footnote{This modification consists just of noting that, by replacing the nonlinear term with distributed delays in  \eqref{Eq_DDE} by $F(x(t), x(t-\tau))$, the identity \eqref{GalerkinCalc_part2a} becomes 
\be
\Bigl \langle \Pi_N \mathcal{F} \Bigl ( \sum_{n=0}^{N-1} y_n(t) \mathcal{K}^\tau_n \Bigr ), \mathcal{K}^\tau_j \Bigr \rangle_{\mathcal{H}} = F \Biggl( \sum_{n=0}^{N-1} y_n(t), \sum_{n=0}^{N-1} y_n(t) K_n(-1) \Biggr);
\ee
{\mg the latter, in turn,} leads to the corresponding change in the formula \eqref{eq:G} for the nonlinear vector field in Eq.~\eqref{Galerkin_cptForm}.
\label{footnote_discrete_delay}}

As shown in the left panel of Figure \ref{Super_chafee} in natural delayed coordinates, the strange attractor associated with Eq.~\eqref{Eq_DDE_chafee} (black) has a nearly symmetric topological shape and is constituted by two fairly high-density``islands'' connected by a foliation of heteroclinic-like orbits. These properties of the attractor are very well
captured by a 6D-Galerkin approximation  (right panel, red).   

These two high-density islands, along with the lower-density areas of heteroclinic-like connections {\mg between the two,} give rise to a bimodal chaotic behavior, which gives rise, in turn, to an interesting time variability. Figure ~\ref{PSD_chafee} 
plots the results of a standard numerical estimation of the spectral density --- also known in the engineering literature as the power spectrum ---  associated with  $x(t)$, as directly simulated using Eq.~\eqref{Eq_DDE_chafee} (black curve) and as approximated by a 6D-Galerkin
approximation (red curve). In both cases, the power spectra are estimated from the corresponding autocorrelation functions\cite{eckmann_ruelle,Ghil2002}.

The numerical results show that the spectrum of $x(t)$ contains two broadband peaks at low frequencies that stand out above an exponentially decaying background. As shown in Fig.~\ref{PSD_chafee}, these
two peaks, as well as the exponential background, are strikingly well approximated by a 6D-Galerkin approximation, in
 both amplitude and frequency, as well as in the rate of decay for high frequencies.

The approximation by a truly low-dimensional, 6D-Galerkin model of these key features of the power spectrum of the solutions to the DDE~\eqref{Eq_DDE_chafee} has deep dynamical implications in terms of the Ruelle-Pollicott resonances and mixing properties of the dynamics on the attractor \cite{Chek_al14_RP}.  These implications are beyond the scope of this article but {\mg we intend to discuss them} elsewhere.  

\br \label{Rmk_Chafee} 
Equation.~\eqref{Eq_DDE_chafee} can actually be seen as a highly idealized ENSO model with memory effects; see \cite{BH89,Cane_al90,Dijkstra05,GCStep15,Galanti_al00, GZT08, Sieber2014, Munnich_al91, TSCJ94,Zaliapin_al10,Zivkovic_al13} and references therein. Indeed, for the given parameter values, the solution of  Eq.~\eqref{Eq_DDE_chafee} admits two metastable states, as can be seen from the two islands in the attractor given in Figure~\ref{Super_chafee}. These two metastable states {\mg are analogous to the warm, El Ni\~no and the cold, La Ni\~na} states in ENSO dynamics.  Moreover, the two broadband peaks at low frequencies in the 
spectral density of the solution, as shown in Figure~\ref{PSD_chafee}, are also reminiscent of the quasi-quadrennial and the quasi-biennial mode in ENSO \cite{MG_AWR'00, Ghil2002, Jiang95} dynamics. The important role of such low-frequency variability in the understanding and prediction of climate dynamics on various time scales was emphasized in \cite{CKG11,HD_MG'05, Ghil2001, MG_AWR'00, MG_AWR'02} and references therein.
\qed

\er

%%%%%%%%%%%%%%%%%%%%%%%%%%
\subsection{A highly idealized ENSO model with memory effects} \label{Sect_ENSO}

In this section, we consider the following {\mg periodically} forced DDE with two discrete delays \cite{GZT08}:
\be \label{ENSO_model}
\dot{x}= -\alpha \tanh(\kappa x(t-\tau_1)) + \beta \tanh(\kappa x(t-\tau_2))  + \gamma \cos(2\pi t).
\ee
This equation is a slightly modified version of the model used in \cite{TSCJ94} for the study of the interaction between the seasonal forcing and the intrinsic ENSO variability.\footnote{The nonlinearity used in \cite{TSCJ94} consists of a sigmoid made of two $\tanh$ segments joined continuously by a line segment, cf. \cite[Eq.~(9)]{Munnich_al91}; this sigmoid is simplified here to be just a hyperbolic tangent function.}  We also refer to \cite{BH89,Cane_al90,Dijkstra05,GCStep15,Galanti_al00,GZT08, Munnich_al91,TSCJ94,Zaliapin_al10,Zivkovic_al13,GZ15} and references therein for other models with retarded arguments used in this context.

The purpose of this subsection is to 
show that the Galerkin scheme developed in this article can also be easily adapted to deal with DDEs with multiple delays, as well as with non-autonomous, forced DDEs.  For this purpose, Eq.~\eqref{ENSO_model} is placed in a quasi-periodic regime {\mg by choosing the parameter values $ \alpha = 2.1, \beta = 1.05, \gamma =3, \kappa = 10, \tau_1 = 0.95,$ and $\tau_2 = 5.13$.}

We outline now the necessary modifications to the nonlinear term $G(y)$ in the Galerkin system \eqref{Galerkin_cptForm} for the case of multiple discrete delays. As a direct generalization of the case with a single discrete delay, given in footnote~\ref{footnote_discrete_delay}, the nonlinearity $F$ in  \eqref{Eq_DDE} takes the form $F(x(t), x(t-\tau_1), \cdots, x(t-\tau_p))$, where $0 < \tau_1 < \cdots < \tau_p=:\tau$.  In this more general situation, the identity \eqref{GalerkinCalc_part2a} becomes
\bea
\hspace*{-1em}\Bigl \langle \Pi_N \mathcal{F} \Bigl ( \sum_{n=0}^{N-1} y_n(t) \mathcal{K}^\tau_n \Bigr ), \mathcal{K}^\tau_j \Bigr \rangle_{\mathcal{H}} & = F \Biggl( \sum_{n=0}^{N-1} y_n(t), \sum_{n=0}^{N-1} y_n(t) K_n(-\frac{\tau_1}{\tau}), \cdots, \\
& \qquad \sum_{n=0}^{N-1} y_n(t) K_n(-\frac{\tau_{p-1}}{\tau}), \sum_{n=0}^{N-1} y_n(t) K_n(-1) \Biggr),
\eea
which leads to the corresponding change in the formula \eqref{eq:G} for the nonlinear vector field in Eq.~\eqref{Galerkin_cptForm}. Note also that the forcing term is dealt with in Remark~\ref{Rmk_Galerkin_forcing}.

\begin{figure}[hbtp]
   \centering
\includegraphics[height=0.4\textwidth,width=.8\textwidth]{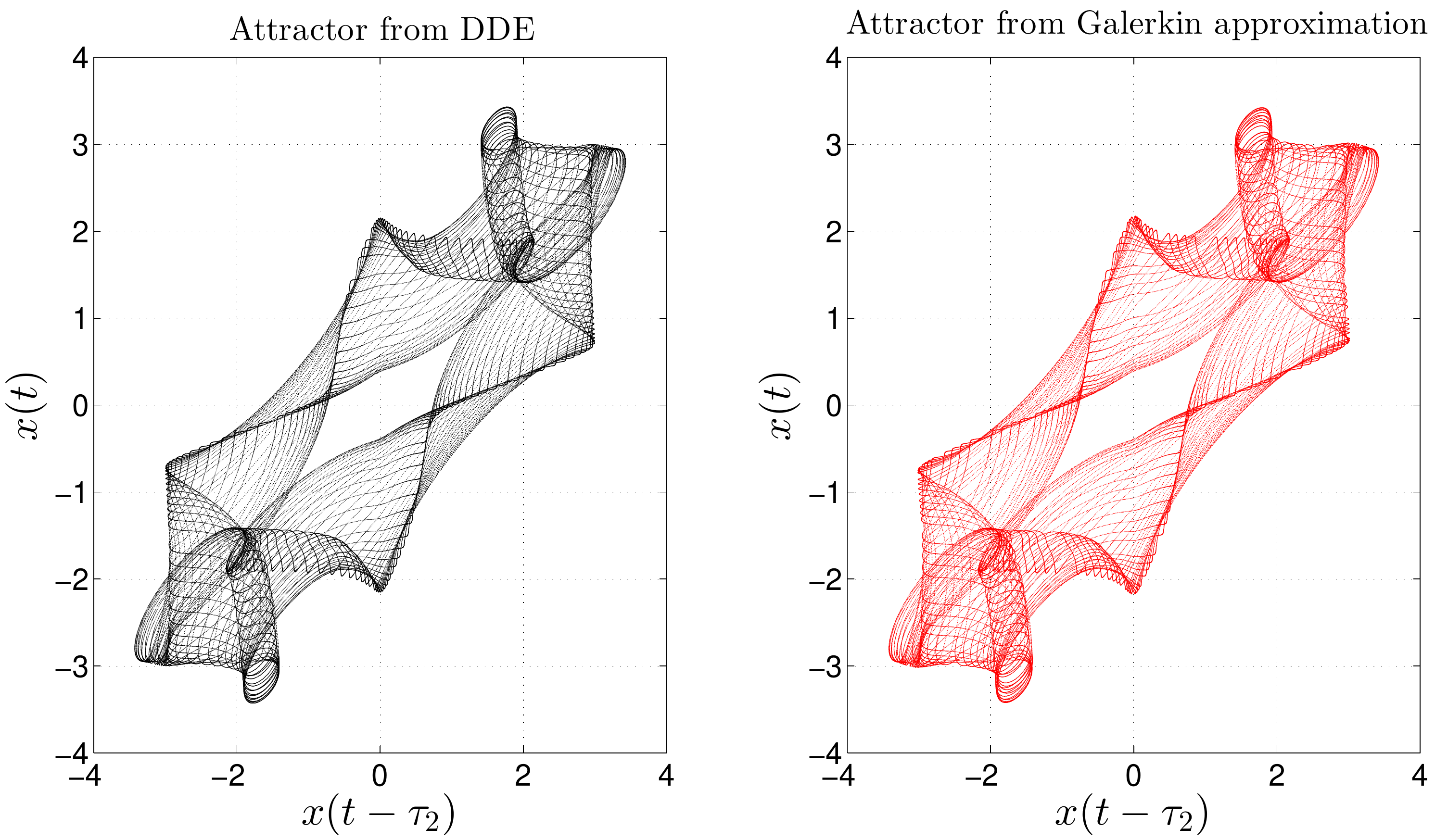}
  \caption{{\footnotesize The attractor associated with Eq.~\eqref{ENSO_model} (left panel) and its approximation obtained from a 40D-Galerkin approximation (right panel).}}   \label{Fig_ENSO_attractors}
\end{figure}

Again, we compare the DDE's attractor in the left panel of Fig.~\ref{Fig_ENSO_attractors} with the attractor obtained from an associated Galerkin system,  in the right panel.  Despite the complexity of the DDE's attractor, a $40$-dimensional Galerkin system can already provide a very accurate reconstruction of this attractor. The need for a higher dimensionality of the Galerkin approximation in this case arises from the presence of incommensurable frequencies in the periodically forced model, namely the seasonal cycle and the internal frequencies \cite{Dijkstra05, JNG'94, MG_AWR'00, MG_AWR'02, TSCJ94}.

%%%%%%%%%%%%%%%%%%%%%%%%%%

\section*{Acknowledgments} 

This work has been partially supported by the Office of Naval Research (ONR) Multidisciplinary University Research Initiative (MURI) grant N00014-12-1-0911 (MDC, HL and MG) and by National Science Foundation (NSF) grant DMS-1049253 (SW).

%%%%%%%%
\appendix

\section{Proof of Theorem \ref{thm:Pn}} \label{sect:thm_Pn_proof}

The identity \eqref{eq:Pn2} follows from the definition of $K_n$ given in \eqref{eq:Pn} and the properties of the Legendre polynomials $L_n$ given by \eqref{eq:Ln_recur} and \eqref{eq:dLn}. The normalization $K_n(1) = 1$ results directly from the identity \eqref{eq:Pn2} and the normalization $L_n(1) = 1$ as recalled in \eqref{eq:Ln_normalization}.

The orthogonality property of $(K_n, K_n(1))$ {\mk is proved in  \cite[Theorem 3.1]{Koo84}.} The formula \eqref{eq:Pn_norm} about the norm can be checked directly by using \eqref{eq:Pn2} and the orthogonality property of the Legendre polynomials recalled in \eqref{eq:Ln_orth}. 

 Finally,   it remains to show that $\Big\{\frac{\mathcal{K}_n}{\|\mathcal{K}_n\|_{\mathcal{E}}}\Big\}$ forms a Hilbert basis of  $\mathcal{E}$  
 which due to the orthonormality property boils down to show that the linear space spanned by the $\mathcal{K}_n$'s  is dense in $\mathcal{E}$.  In that respect, by recalling that $\mathcal{K}_n=(K_n,K_n(1))$ and that $K_n$  are polynomials of degree $n$ orthogonal under the inner product $\langle \cdot,\cdot\rangle_{\mathcal{E}}$,  it follows that any function in the subspace 
\bes
\mathcal{S}_n:=\{(f, f(1)) : \text{$f$ is a polynomial of degree $n$}\}
\ees 
admits a unique expansion in terms of $\{(K_0, K_0(1)), \cdots, (K_{n}, K_{n}(1))\}$. Note also that $\bigcup_{n = 0}^\infty \mathcal{S}_n$ is dense in the subspace 
\bes
\mathcal{C}:= \{(f, f(1)) : \text{$f$ is a continuous function on $[-1,1]$}\},
\ees 
which is itself dense in 
\bes
\widetilde{\mathcal{C}}:= \{(f, a) : \text{$f$ is a continuous function on $[-1,1]$, and $a \in \mathbb{R}$}\}.
\ees
Since the last subspace $\widetilde{\mathcal{C}}$ is dense in $\mathcal{E}$, the desired result follows.
 \qed

\section{Proof of Proposition \ref{prop:dPn}} \label{sect:coef_matrix_proof}
Note that $a_n := (a_{n,0}, \cdots, a_{n,n-1})^\tr$ given in \eqref{eq:dPn} can be obtained by rewriting both sides in terms of the Legendre polynomials and comparing the coefficients, which leads to the algebraic equation $\mathbf{T} a_n = b_n$ to be satisfied by $a_n$. We provide below some details about the derivation of the matrix $\mathbf{T}$ and the vector $b$ given in \eqref{eq:algebraic_def}.

First note that by the definition of $K_n$ given in \eqref{eq:Pn}, we get
\be
\frac{\d K_n}{\d s}(s)  = \frac{\d}{\d s}\left[-(1+s)\frac{\d}{\d s} L_n(s) \right] + ( n^2 + n + 1) \frac{\d L_n(s)}{\d s}.
\ee
By using then the recurrence formula for the derivative of $L_n$ given in \eqref{eq:dLn}, we obtain
\bea \label{eq:LHS_dPn}
\frac{\d K_n}{\d s}(s)  & = \frac{\d}{\d s}\left[-(1+s) \sum_{k \in I_n} (2k+1)L_k(s) \right] \\
& \qquad  + ( n^2 + n + 1) \sum_{k \in I_n} (2k+1)L_k(s) \\
& = - \sum_{k \in I_n} (2k+1)L_k(s) - (1+s) \sum_{k \in I_n} (2k+1) \frac{\d L_k(s)}{\d s} \\
& \qquad + ( n^2 + n + 1) \sum_{k \in I_n} (2k+1)L_k(s) \\
& = ( n^2 + n) \sum_{k \in I_n} (2k+1)L_k(s) \\
& \qquad - (1+s) \sum_{k \in I_n} \sum_{j \in I_k} (2k+1) (2j+1)L_j(s),
\eea
where we recall from \eqref{eq:idx_n} that $I_n =\{k \in \mathbb{N} : 0\le k \le n-1, k+n \text{ is odd}\}$.

By using again the definition of $K_n$ given in \eqref{eq:Pn}, the RHS of \eqref{eq:dPn} can be rewritten in terms of the Legendre polynomials $L_n$ as follows:
\bea \label{eq:RHS_dPn}
& \sum_{k  = 0}^{n-1} a_{n,k} K_k(s) \\
& = \sum_{k  = 0}^{n-1} a_{n,k} \left[ -(1+s)\frac{\d}{\d s} L_k(s) + ( k^2 + k + 1)  L_k(s) \right ] \\
& =  -(1+s) \sum_{k  = 0}^{n-1} \sum_{j \in I_k} a_{n,k}(2j+1)L_j(s)  +  \sum_{k  = 0}^{n-1} a_{n,k} ( k^2 + k + 1)  L_k(s).
\eea
Now, by using \eqref{eq:LHS_dPn} and \eqref{eq:RHS_dPn} in the expansion $\frac{\d K_n}{\d s}(s) = \sum_{k  = 0}^{n-1} a_{n,k} K_k(s)$, we obtain
\bea \label{eq:dPn_rewrite}
 & - (1+s) \sum_{k  = 0}^{n-1} \sum_{j \in I_k} a_{n,k}(2j+1)L_j(s)  +  \sum_{k  = 0}^{n-1} a_{n,k} ( k^2 + k + 1)  L_k(s) \\
& = ( n^2 + n ) \sum_{k \in I_n} (2k+1)L_k(s) - (1+s) \sum_{k \in I_n} \sum_{j \in I_k} (2k+1) (2j+1)L_j(s).
\eea
Finally, by using the three-term recurrence formula $ (j+1) L_{j+1}(s)  = (2j+1) s L_j(s) - j L_{j-1}(s)$ recalled in \eqref{eq:Ln_recur}, we obtain from Eq.~\eqref{eq:dPn_rewrite} that
\bea
 & - \sum_{k  = 0}^{n-1} \sum_{j \in I_k} a_{n,k} \left[ (j+1) L_{j+1}(s) + (2j+1) L_j(s) +  j L_{j-1}(s) \right ] \\
& + \sum_{k  = 0}^{n-1} a_{n,k} ( k^2 + k + 1)  L_k(s)  = ( n^2 + n) \sum_{k \in I_n} (2k+1)L_k(s)  \\
&  - \sum_{k \in I_n} \sum_{j \in I_k} (2k+1) \left[ (j+1) L_{j+1}(s) + (2j+1)L_j(s) +  j L_{j-1}(s) \right ].
\eea
The system of algebraic equations given in \eqref{eq:algebraic}--\eqref{eq:algebraic_def} can then be obtained by equating the coefficients for $L_{n-1}, \cdots, L_0$ on both sides of the last equation above. The  proof is now complete. \qed

%\epp

%%%%%%%%%%%%%%%%%%%%%%%%%%%%%%
\section{Analytic formulas for nonlinear systems of DDEs}\label{Appendix_systems}

In this section, we extend the analytic formulas of Section~\ref{Sec_Galerkin_analytic} to nonlinear systems of delay equations of the form \eqref{Eq_nln_sys} based on the vectorized Koornwinder polynomials introduced in subsection~\ref{Sec_Vectorization}.

Recall that \eqref{Eq_nln_sys}  is given by
\be \label{Eq_DDE_system_recall}
\frac{\d \mathbf{x}}{\d t}=L_S \mathbf{x}(t)+ B \mathbf{x}(t-\tau) +  \int_{t-\tau}^t  C(s-t) \mathbf{x}(s) \d s + \mathbf{F}\Big(\mathbf{x}(t),\int_{t-\tau}^t  \mathbf{x}(s) \d s\Big),
\ee
where $\mathbf{F}:\mathbb{R}^d\times \mathbb{R}^d  \rightarrow \mathbb{R}^d$, and $L_S$, $B$, and $C$ are as given in \eqref{Def_LD}.

Recall also that the unknown function $u_N$ in the corresponding Galerkin system \eqref{Eq_DDE_Galerkin} takes values in the $Nd$-dimensional subspace $\mathcal{H}_{N}$ defined in \eqref{subspace_HNd}.  We can thus rewrite $u_N$ in terms of the first $Nd$ vectorized Koornwinder polynomials (see \eqref{Eq_superKoor}) as follows: 
\be \label{x_t expand_system}
u_N(t) = \sum_{j=1}^{Nd} y_j(t) \mathbb{K}^\tau_j, \qquad t \ge 0,
\ee
where 
\be \label{Galerkin_comp_system_case}
y_j(t) = \frac{\langle u_N(t),  \mathbb{K}^\tau_j  \rangle_{\mathcal{H}}}{\|  \mathcal{K}^\tau_{j_q} \|^2_{\mathcal{H}_1}}, \qquad j = 1, \cdots, Nd,
\ee
with $j_q$ determined by \eqref{index_relation} and $\mathcal{H}_1$ being the product space $L^2([-\tau,0); \mathbb{R})\times \mathbb{R}$ equipped with the inner product given by \eqref{H_inner}.

In the following, we first deal with a special case in subsection~\ref{Subsect_const_C} by assuming that the linear operator $C$ in the system \eqref{Eq_DDE_system_recall} is time-independent. Necessary changes for the time-dependent case is then outlined in subsection~\ref{Subsect_var_C}.

%%%%%%%%%%%%%%%%
\subsection{The case when $C$ in \eqref{Eq_DDE_system_recall}  is time-independent} \label{Subsect_const_C}
  
In this case, all of the three linear operators $L_S$, $B$ and $C$ are $d\times d$ matrices. By the same type of computation as given in Section~\ref{Sec_Galerkin_analytic} for the scalar case, we obtain the following $Nd$-dimensional ODE system for $(y_1, \cdots, y_{Nd})$:
\begin{equation} \label{Galerkin_AnalForm_for_system}
\begin{aligned}
\frac{\d y_{j}}{\d t} & = \frac{1}{\|\mathcal{K}_{j_q}\|_\mathcal{E}^2 }  
\sum_{n=1}^{Nd} \Biggl[ 
\frac{2}{\tau} \sum_{k=0}^{n_q-1} a_{n_q,k} \left( \delta_{j_q,k} \|\mathcal{K}_{j_q}\|^2_{\mathcal{E}} - 1 \right ) \delta_{n_r, j_r} \\
& \hspace{3em} +  (L_S)_{j_r, n_r} +  K_{n_q}(-1) B_{j_r, n_r}
+ \tau (2 \delta_{n_q, 0} - 1) C_{j_r, n_r} \Biggr ] y_{n}(t) \\
& \hspace{3em} +\frac{1}{\|\mathcal{K}_{j_q}\|_\mathcal{E}^2 }  F_{j_r} \Biggl(
\sum_{n=1}^{Nd} y_{n}(t) \mathbf{K}_n(1), \; 
\tau \sum_{n=1}^{Nd} (2 \delta_{n_q, 0} -1) \mathbf{K}_n(1) y_{n}(t) \Biggr),
\end{aligned}
\end{equation}
where $j = 1, \cdots, Nd$, and where given an integer $l \in \{1, \cdots, Nd\}$, the associated integers $l_q \in \{0, \cdots, N-1\}$ and $l_r \in \{1, \cdots, d\}$  are determined by \eqref{index_relation}-\eqref{index_relationb}. The coefficients $a_{i,j}$ are  determined by \eqref{eq:dPn}, and $\mathcal{E}$ denotes  the product space $L^2([-1,1); \mathbb{R}) \times  \mathbb{R}$ equipped with the inner product given by \eqref{eq:inner_E}.

The above Galerkin system can be put into the following compact form:
\be \label{Galerkin_cptForm_for_system}
\boxed{\frac{\d \boldsymbol{y}}{\d t} = A \boldsymbol{y}  +  G (\boldsymbol{y}),}
\ee
where $A \boldsymbol{y}$ denotes the linear part of  Eq.~\eqref{Galerkin_AnalForm_for_system}, and $G(\boldsymbol{y})$ the nonlinear part. Namely, $A$ is the $Nd\times Nd$ matrix whose elements are given by 
\begin{equation} \label{eq:A_for_system}
\boxed{
\begin{aligned}
A_{j,n} & =  
\frac{1}{\|\mathcal{K}_{j_q}\|_\mathcal{E}^2 }  
\Biggl[ 
\frac{2}{\tau} \sum_{k=0}^{n_q-1} a_{n_q,k} \left( \delta_{j_q,k} \|\mathcal{K}_{j_q}\|^2_{\mathcal{E}} - 1 \right ) \delta_{n_r, j_r} \\
& \hspace{3em} +  (L_S)_{j_r, n_r} +  K_{n_q}(-1) B_{j_r, n_r}
+ \tau (2 \delta_{n_q, 0} - 1) C_{j_r, n_r} \Biggr ],
\end{aligned}
} 
\end{equation}
where $j, n  = 1, \cdots, Nd$, and the nonlinear vector field $G \colon \mathbb{R}^{Nd} \rightarrow \mathbb{R}^{Nd}$, is given component-wisely  
by  
\be \label{eq:G_for_system}
\boxed{G_j(\boldsymbol{y}) = \frac{1}{\|\mathcal{K}_{j_q}\|_\mathcal{E}^2 }  F_{j_r} \Biggl(
\sum_{n=1}^{Nd} y_{n}(t) \mathbf{K}_n(1), \; 
\tau \sum_{n=1}^{Nd} (2 \delta_{n_q, 0} -1) \mathbf{K}_n(1) y_{n}(t) \Biggr).}
\ee

%%%%%%%%%%%%%%%%
\subsection{The case when $C$  in \eqref{Eq_DDE_system_recall}  is time-dependent}
\label{Subsect_var_C}

We outline below necessary modifications to \eqref{eq:A_for_system} when $C \in L^2([-\tau, 0); \mathbb{R}^{d \times d})$. 

Note  that the contribution from the term $\int_{t-\tau}^t  C(s-t) \mathbf{x}(s) \d s$ to the component $A_{j,n}$ given in  \eqref{eq:A_for_system} is: 
\be
\alpha_{j,n} = \Big\langle \int_{-\tau}^0 C(\theta) \mathbf{K}_n^\tau(\theta) \d \theta,  \mathbf{K}_j^\tau(0) \Big\rangle.
\ee
Note also that $\int_{-\tau}^0 C(\theta) \mathbf{K}_n^\tau(\theta) \d \theta$ is a $d$-dimensional column vector, whose $k^{th}$ component is given by $\int_{-\tau}^0 C_{k,n_r}(\theta) K_{n_q}^\tau(\theta) \d \theta$ where $1 \le k \le d$.  

It follows then 
\be
\alpha_{j,n} =  \int_{-\tau}^0 C_{j_r,n_r}(\theta) K_{n_q}^\tau(\theta) \d \theta.
\ee

Since each component of $C$ belongs to $L^2([-\tau, 0); \mathbb{R})$, by the identity \eqref{L2_decomp_0} given in Lemma~\ref{Fundamental_lemma}, we have
\be \label{Expansion_C}
C_{i,j}  = \sum_{l=0}^\infty \frac{\langle C_{i,j}, K_l^\tau \rangle_{L^2} }{\tau \|\mathcal{K}_l^\tau \|_{\mathcal{H}_1}^2 }  K_l^\tau, \quad 1 \le i,j \le d.
\ee
It follows  then
\bes
\alpha_{j,n} = \sum_{l=0}^\infty \frac{\langle C_{j_r,n_r}, K_l^\tau \rangle_{L^2} }{\tau \|\mathcal{K}_l^\tau \|_{\mathcal{H}_1}^2 }   \int_{-\tau}^0 K_l^\tau(\theta) K_{n_q}^\tau(\theta) \d \theta.
\ees
Since $ \langle  K_i^\tau,  K_j^\tau \rangle_{\mathcal{H}_1} = \delta_{i,j}$, we get
\bes
\frac{1}{\tau} \int_{-\tau}^0 K_i^\tau(\theta) K_{j}^\tau(\theta) \d \theta = (\delta_{i,j} \|\mathcal{K}_{i}^\tau\|_{\mathcal{H}_1}^2 - 1), \quad i,j \in \mathbb{N},
\ees
leading thus to 
\bea
\alpha_{j,n} & = \sum_{l=0}^\infty \frac{\langle C_{j_r,n_r}, K_l^\tau \rangle_{L^2} }{\|\mathcal{K}_l^\tau \|_{\mathcal{H}_1}^2 }  (\delta_{l,n_q} \|\mathcal{K}_{l}^\tau\|_{\mathcal{H}_1}^2 - 1) \\
& = \langle C_{j_r,n_r}, K_{n_q}^\tau \rangle_{L^2}  - \sum_{l=0}^\infty \frac{\langle C_{j_r,n_r}, K_l^\tau \rangle_{L^2} }{\|\mathcal{K}_l^\tau \|_{\mathcal{H}_1}^2 } \\
& = \langle C_{j_r,n_r}, K_{n_q}^\tau \rangle_{L^2},
\eea
where we have used $\sum_{l=0}^\infty \frac{\langle C_{j_r,n_r}, K_l^\tau \rangle_{L^2} }{\|\mathcal{K}_l^\tau \|_{\mathcal{H}_1}^2 } = 0$, which follows from the identity \eqref{Eq_identity1_0} in Lemma~\ref{Fundamental_lemma}. 

So the only modification to \eqref{eq:A_for_system} is to replace the term $\tau (2 \delta_{n_q, 0} - 1) C_{j_r, n_r}$ by $\langle C_{j_r,n_r}, K_{n_q}^\tau \rangle_{L^2}$. Note that when $C_{j_r,n_r} \in L^2([-\tau, 0); \mathbb{R})$ is a constant function, the latter is reduced into the former since
\bes 
\int_{-\tau}^0 K^\tau_n(\theta) \d \theta  = \tau (2 \delta_{n,0} - 1), \quad n \in \mathbb{N}.
\ees

To summarize, when $C$ is time-dependent, the $Nd$-dimensional Galerkin approximation associated with \eqref{Eq_DDE_system_recall} still takes the form \eqref{Galerkin_cptForm_for_system} with the nonlinear part $G(\boldsymbol{y})$  given by \eqref{eq:G_for_system}, and the elements of the $Nd\times Nd$ matrix $A$, given by
\begin{equation} \label{eq:A_for_system_var_C}
\boxed{
\begin{aligned}
A_{j,n} & =  
\frac{1}{\|\mathcal{K}_{j_q}\|_\mathcal{E}^2 }  
\Biggl[ 
\frac{2}{\tau} \sum_{k=0}^{n_q-1} a_{n_q,k} \left( \delta_{j_q,k} \|\mathcal{K}_{j_q}\|^2_{\mathcal{E}} - 1 \right ) \delta_{n_r, j_r} \\
& \hspace{3em} +  (L_S)_{j_r, n_r} +  K_{n_q}(-1) B_{j_r, n_r}
+ \langle C_{j_r,n_r}, K_{n_q}^\tau \rangle_{L^2} \Biggr ].
\end{aligned}
} 
\end{equation}

%%%%%%%%%%%%%%%%%%
\bibliographystyle{amsalpha}
%\bibliography{DDE_bib-MG}
\newcommand{\etalchar}[1]{$^{#1}$}
\providecommand{\bysame}{\leavevmode\hbox to3em{\hrulefill}\thinspace}
\providecommand{\MR}{\relax\ifhmode\unskip\space\fi MR }
\providecommand{\MRhref}[2]{%
  \href{http://www.ams.org/mathscinet-getitem?mr=#1}{#2}
}
\providecommand{\href}[2]{#2}

\end{document}